\documentclass{jfm}

\usepackage{amsmath,amsfonts,latexsym}
\usepackage{graphicx}
\usepackage{subfigure}
\usepackage{bm}
\usepackage{multirow}
\usepackage{array}
\usepackage{setspace}
\usepackage{xcolor}
\usepackage{natbib}
\usepackage{pxfonts}
\usepackage{caption}

\ifCUPmtlplainloaded \else
  \checkfont{eurm10}
  \iffontfound
    \IfFileExists{upmath.sty}
      {\typeout{^^JFound AMS Euler Roman fonts on the system,
                   using the 'upmath' package.^^J}%
       \usepackage{upmath}}
      {\typeout{^^JFound AMS Euler Roman fonts on the system, but you
                   dont seem to have the}%
       \typeout{'upmath' package installed. JFM.cls can take advantage
                 of these fonts,^^Jif you use 'upmath' package.^^J}%
      }
  \else
  \fi
\fi

\ifCUPmtlplainloaded \else
  \checkfont{msam10}
  \iffontfound
    \IfFileExists{amssymb.sty}
      {\typeout{^^JFound AMS Symbol fonts on the system, using the
                'amssymb' package.^^J}%
       \usepackage{amssymb}%
       \let\le=\leqslant  
       \let\ge=\geqslant  
      }{}
  \fi
\fi

\ifCUPmtlplainloaded \else
  \IfFileExists{amsbsy.sty}
    {\typeout{^^JFound the 'amsbsy' package on the system, using it.^^J}%
     \usepackage{amsbsy}}
    {}
\fi

\newsavebox{\astrutbox}
\sbox{\astrutbox}{\rule[-5pt]{0pt}{20pt}}

\title[Long ring waves in a stratified fluid over a shear flow]{Long ring waves in a stratified fluid\\ over a shear flow}

\author[K.R. Khusnutdinova, X. Zhang]%
{K. R.\ns K\ls H\ls U\ls S\ls N\ls U\ls T\ls D\ls I\ls N\ls O\ls V\ls A%
	\thanks{Email address for correspondence: K.Khusnutdinova@lboro.ac.uk}, \and 
X.\ns Z\ls H\ls A\ls N\ls G}

\affiliation{Department of Mathematical Sciences, Loughborough University, Loughborough LE11 3TU, UK}

\date{?; revised ?; accepted ?. - To be entered by editorial office}

\begin{document}

\maketitle

\begin{abstract}
Oceanic waves registered by satellite observations often have curvilinear fronts and propagate over various currents. In this paper, we study long linear and weakly-nonlinear ring waves in a stratified fluid in the presence of a depth-dependent horizontal shear flow.  It is shown that despite the clashing geometries of the waves and the shear flow, there exists a linear modal decomposition (different from the known decomposition in Cartesian geometry), which can be used to describe distortion of the wavefronts of surface and internal waves, and systematically derive a 2+1 - dimensional cylindrical Korteweg-de Vries - type equation for the amplitudes of the waves. The general theory is applied to the case of the waves in a two-layer fluid with a piecewise-constant current, with an emphasis on the effect of the shear flow on the geometry of the wavefronts. The distortion of the wavefronts is described by the singular solution (envelope of the general solution) of the nonlinear first-order differential equation, constituting generalisation  of the dispersion relation in this curvilinear geometry.  There exists a striking difference in the shape of the wavefronts of surface and interfacial waves propagating over the same shear flow.
\end{abstract}





\section{Introduction}

The Korteweg-de Vries (KdV) equation and its generalisations are successfully used to describe
long weakly-nonlinear internal waves that are commonly observed in the oceans  \citep{Benjamin66, Benney66, MR, Grimshaw01, Helfrich06, Grimshaw97, GOSS98, Grue, Apel07}, as well as describing weakly-nonlinear shallow-water surface waves \citep{Boussinesq, KdV, Ablowitz, Kodama}.  The waves described by these models have plane or nearly-plane fronts. However, waves registered by satellite observations often look like a part of a ring,  motivating the present study  of annular waves (see Figure 1 for the image of a nearly annular internal soliton found at $http://www.lpi.usra.edu/publications/slidesets/oceans/oceanviews/oceanviews\_index.shtml$).  A very large collection of images of various nonlinear internal waves can be found at \linebreak $ http://www.internalwaveatlas.com/Atlas2\_index.html. $ Observations of internal waves with curvilinear fronts have been reported in several studies, e.g. \cite{Farmer, Nash, Jackson}. Internal waves have a strong effect on acoustic signalling, as well  impacting submersibles, offshore structures, and underwater pipelines. They also significantly contribute to the ocean mixing processes. Therefore, it is important to develop a good understanding of the main properties and behaviour of the waves. In natural oceanic environments these waves often propagate over currents, for example, tides, wind-drift currents, river flows, etc. Thus, we aim to develop an asymptotic theory which could be used to model long ring waves in a stratified fluid over a shear flow. In this paper we focus on the basic balance between nonlinearity and dispersion, considering waves with cylindrical divergence on horizontal shear flows in the KdV regime.

\begin{figure}
           \centering\includegraphics[width=0.8\columnwidth]{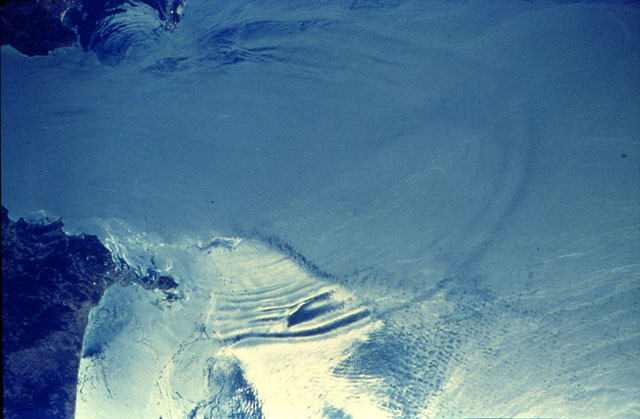} 
		\caption {Internal soliton generated in the Strait of Gibraltar (NASA image STS17-34-098, Lunar and Planetary Institute).}
\end{figure}

Currently, there is a considerable interest to both internal and surface waves on sheared currents. Among recent results, the effect of a shear flow on the linear surface ship waves was studied by \cite{Ellingsen}, while approximations to periodic travelling surface wave profiles in flows with constant vorticity have been obtained by \cite{Constantin}.   \cite{Thomas} have studied how vorticity modifies the modulational instability properties of weakly-nonlinear surface wave trains. Analytical considerations of the latter paper complemented an earlier study by \cite{Oikawa} and a strongly-nonlinear numerical analysis developed by \cite{Nwogu}.

The complementary nature of the analytical weakly-nonlinear studies and strongly-nonlinear numerical modelling is also notable in the research on internal waves. Analytical studies help to identify the main trends and dependencies, creating the framework for the discussion of the observed and modelled phenomena, while advanced numerical modelling allows one to describe important features of the waves in more realistic settings not amenable to theoretical analysis (see, for example, the reviews by \cite{Helfrich06,Apel07}).

The effects of various sheared currents on internal waves and their role in the oceanic processes have been studied by a number of authors (e.g.,  \cite{Lee,Mooers, Olbers, Young, Voronovich, Buhler}). Modifications of large internal solitary waves by a background shear flow have been studied analytically by \cite{Choi} and modelled numerically, for example, by \cite{Stastna1} and \cite{Stastna2}, within the framework of two - dimensional simulations. Three - dimensional numerical modelling, although expensive, has also been developed. In particular, large-amplitude internal waves in the Strait of Gibraltar have been modelled by \cite{Vlasenko1} and \cite{Sannino}. Tidal generation of internal waves in the Celtic Sea has been modelled by \cite{Vlasenko2,Vlasenko3} and \cite{Grue15}. We note that the waves generated  in the Strait of Gibraltar are nearly annular (see Figure 1), as are the waves generated near a sea mountain in the Celtic Sea (see Figure A1 in \cite{Vlasenko3}).

The cylindrical (or concentric) Korteweg-de Vries (cKdV) equation is a universal weakly-nonlinear weakly-dispersive wave equation in cylindrical geometry. Originally derived in the context of ion-acoustic waves in plasma \citep{Maxon74}, it was later derived for the surface waves in a uniform fluid, first from Boussinesq equations \citep{Miles78},  and then from the full Euler equations  \citep{Johnson80}. Versions of the equation were also derived for internal waves in a stratified fluid without shear flow \citep{Lipovskii85}, and surface waves in a uniform fluid with a shear flow \citep{Johnson90}. The original equation is integrable  \citep{Calogero}, and there exists a useful map between cKdV and KP equations \citep{Johnson80,Johnson_book,Klein07}, while a generic shear flow leads to a nonintegrable cKdV - type equation.

In this paper, we study the propagation of internal and surface ring waves in a stratified fluid over a prescribed shear flow, generalising the previous studies. The paper is organised as follows. In section 2 we derive a 2+1 - dimensional cKdV-type model for the amplitudes of surface and internal waves, by finding an appropriate linear modal decomposition (different from the known modal decomposition in Cartesian coordinates) and using techniques from the asymptotic multiple - scales analysis. The detailed structure of the wavefronts is analysed analytically, for the case of a two-layer model with a piecewise-constant current, in section 3.  We also obtain conditions guaranteeing that there are no critical levels, and calculate explicit expressions for the coefficients of the derived amplitude equation, listed in Appendix A.  Some conclusions are drawn in section 4.


 \section{Problem formulation and amplitude equation}

We consider a ring wave propagating in an inviscid incompressible fluid, described by the full set of Euler equations:
\begin{eqnarray}
&& \rho (u_t + u u_x + v u_y + w u_z) + p_x = 0, \label{1} \\
&& \rho (v_t + u v_x + v v_y + w v_z) + p_y = 0, \label{2} \\
&& \rho (w_t + u w_x + v w_y + w w_z) + p_z + \rho g = 0, \label{3} \\
&& \rho_t + u \rho_x + v \rho_y + w \rho_z = 0, \label{4} \\
&& u_x + v_y + w_z = 0, \label{5} 
\end{eqnarray}
with the free surface and rigid bottom boundary conditions appropriate for the oceanic conditions:
\begin{eqnarray}
&&w = h_t + u h_x + v h_y \quad \mbox{at} \quad z = h(x,y,t), \label{6} \\
&&p = p_a \quad \mbox{at} \quad z = h(x,y,t), \label{7} \\
&&w = 0 \quad \mbox{at} \quad z = 0. \label{8}
\end{eqnarray}
Here, $u,v,w$ are the velocity components in $x,y,z$ directions respectively, $p$ is the pressure, $\rho$ is the density, $g$ is the gravitational acceleration, $z = h(x,y,t)$ is the free surface depth (with $z = 0$ at the bottom), and $p_a$ is the constant atmospheric pressure at the surface. We assume that in the basic state $u_0 = u_0(z), ~ v_0 = w_0 = 0, ~p_{0z} = - \rho_0 g, ~h = h_0$. Here $u_0(z)$ is a horizontal  shear flow in the $x$-direction, and  $\rho_0 = \rho_0(z)$ is a stable background density stratification (see Figure 2).

\begin{figure}
		\centering \includegraphics[width=0.7\columnwidth]{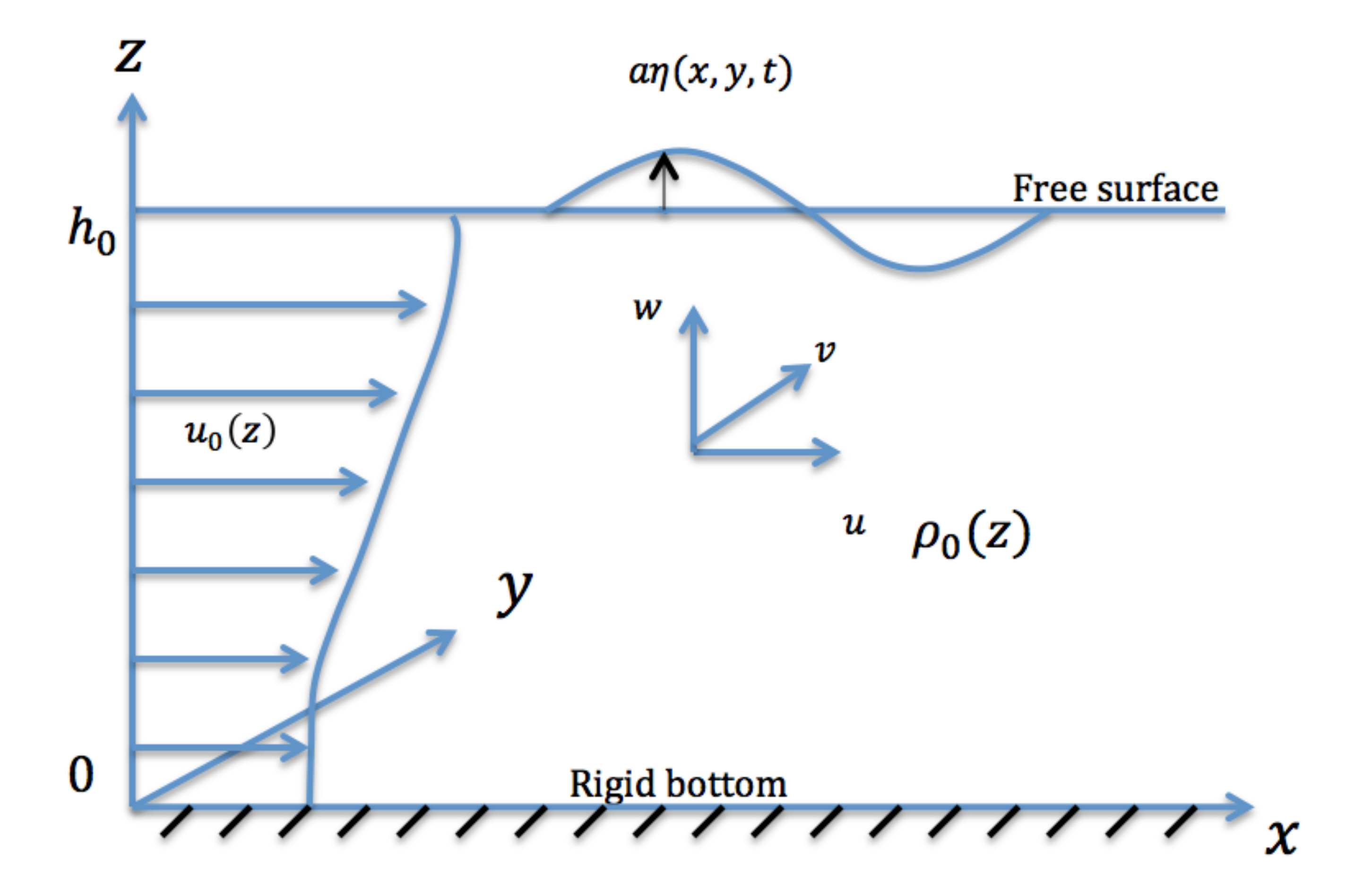} 
		\caption {Schematic of the problem formulation.} 
\end{figure}

	It is convenient to use the vertical particle displacement $\zeta$ as an additional dependent variable, which is defined by the equation
\begin{equation}
\zeta_t + u \zeta_x + v \zeta_y + w \zeta_z = w, \label{9}
\end{equation}
and satisfies the obvious surface boundary condition 
\begin{equation}
\zeta = h - h_0, \label{10}
\end{equation}
where $h_0$ is the unperturbed depth of the fluid.

We aim to derive an amplitude equation for the amplitudes of the long surface and internal waves. Thus, we use the following non-dimensional variables:
\begin{eqnarray*}
&&x \to \lambda x, \quad y \to \lambda y, \quad z \to h_0 z, \quad t \to \frac{\lambda}{c^*}t, \\
&& u \to c^* u, \quad v \to c^* v, \quad w \to \frac{h_0 c^*}{\lambda} w, \\
&&(\rho_0, \rho) \to \rho^*( \rho_0, \rho),  \quad h \to h_0 + a \eta, \quad p \to p_a + \int_{z}^{h_0} \rho^* \rho_0(s) g ~\mathrm{d} s + \rho^* g h_0 p.
\end{eqnarray*}
Here, $\lambda$ is the wave length, $a$ is the wave amplitude, $c^*$ is the typical long-wave speed of surface or internal waves ($\sqrt{g h_0}$ or $h^* N^*$, respectively, where $N^*$ is a typical value of the buoyancy frequency, and $h^*$ is a typical depth of the stratified layer), $\rho^*$ is the dimensional reference density of the fluid, while $\rho_0(z)$ is the non-dimensional function describing stratification in the basic state, and $\eta = \eta(x,y,t)$ is the non-dimensional free surface elevation. In both cases non-dimensionalisation leads to the appearance of the same small parameters in the problem, the amplitude parameter $\varepsilon = a/h_0$ and the wavelength parameter $\delta = h_0/\lambda$. In the second case, a third small (Boussinesq) parameter will appear as well, but the Boussinesq approximation is not used in the subsequent derivation. Thus, it is natural to non-dimensionalise the general problem formulation, including both surface and internal waves,  using the parameters of the faster surface waves, and measure the speeds of the internal waves as fractions of the surface wave speed, etc. However, if one is primarily interested in the study of internal waves, it is more natural to use the typical speed of the internal waves. 

Introducing the cylindrical coordinate system moving at a constant speed $c$ (a natural choice is the flow speed at the bottom, as we will show later), considering deviations from the basic state (the same notations $u$ and $v$ have been used for the projections of the deviations of the speed on the new coordinate axis), and scaling the appropriate variables using the amplitude parameter $\varepsilon$,
\begin{eqnarray*}
&&x \to ct + r \cos \theta, \quad y \to r \sin \theta, \quad z \to z, \quad t \to t, \\
&&u \to u_0(z) + \varepsilon (u \cos \theta - v \sin \theta), \quad v \to \varepsilon (u \sin \theta + v \cos \theta), \\
&& w \to \varepsilon w, \quad p \to  \varepsilon p, \quad \rho \to \rho_0 + \varepsilon \rho,
 \end{eqnarray*}
 we arrive at the following non-dimensional problem formulation in the moving cylindrical coordinate frame:
 \begin{eqnarray}
 &&(\rho_0 + \varepsilon \rho) \left [u_t + \varepsilon \left (u u_r + \frac{v}{r} u_{\theta} + w u_z  - \frac{v^2}{r}\right )  + ((u_0-c) u_r + u_{0z} w) \cos \theta  \right .\nonumber \\
 && \hspace{6cm} \left .  - (u_0-c) (u_{\theta}-v) \frac{\sin \theta}{r}\right ] + p_r = 0, \label{1a}\\
 &&(\rho_0 + \varepsilon \rho) \left [v_t + \varepsilon \left (u v_r + \frac{v}{r} v_{\theta} + w v_z + \frac{uv}{r}\right )  + (u_0-c) v_r \cos \theta \right .  \nonumber \\
 && \hspace{5cm} \left .  - \left ((u_0-c)(\frac{v_{\theta}}{r} + \frac{u}{r}) + u_{0z} w\right ) \sin \theta \right ] + \frac{p_{\theta}}{r} = 0, \label{2a}\\
 && \delta^2 (\rho_0 + \varepsilon \rho) \left [ w_t + \varepsilon \left (u w_r + \frac{v}{r} w_{\theta} + w w_z \right ) + (u_0-c) \left ( w_r \cos \theta - w_{\theta} \frac{\sin \theta}{r} \right ) \right ] \nonumber  \\
 &&\hspace{9cm} + p_z + \rho = 0, \label{3a}\\
 &&\rho_t + \varepsilon \left (u \rho_r + \frac{v}{r} \rho_{\theta} + w \rho_z \right ) + (u_0-c) \left (\rho_r \cos \theta - \rho_{\theta} \frac{\sin \theta}{r} \right ) + \rho_{0z} w = 0, \label{4a} \\
 &&u_r + \frac{u}{r} + \frac{v_{\theta}}{r} + w_z = 0, \label{5a}\\
 &&w = \eta_t + \varepsilon \left (u \eta_r + \frac{v}{r} \eta_{\theta} \right ) + (u_0-c) \left (\eta_r \cos \theta - \eta_{\theta} \frac{\sin \theta}{r} \right ) \quad \mbox{at} \quad z = 1 + \varepsilon \eta, \label{6a}\\
 &&\varepsilon p = \int_{1}^{1 + \varepsilon \eta} \rho_0(s) ds \quad \mbox{at} \quad z = 1 + \varepsilon \eta, \label{7a} \\
 &&w = 0 \quad \mbox{at} \quad z=0, \label{8a}
 \end{eqnarray}
 with the vertical particle displacement satisfying the following equation and boundary condition:
 \begin{eqnarray}
 &&\zeta_t + \varepsilon \left (u \zeta_r + \frac{v}{r} \zeta_{\theta} + w \zeta_z\right ) + (u_0-c) \left ( \zeta_r \cos \theta - \zeta_{\theta} \frac{\sin \theta}{r} \right ) = w, \label{9a}\\
 &&\zeta = \eta \quad \mbox{at} \quad z = 1 + \varepsilon \eta. \label{10a}
 \end{eqnarray}
 For the sake of simplicity, in the subsequent derivation we impose the condition $\delta^2 = \varepsilon$, although this is not the necessary condition. Indeed, variables can be scaled further to replace $\delta^2$ with $\varepsilon$ in the equations  \citep{Johnson_book}.  

 We look for a solution of this problem in the form of asymptotic multiple - scales expansions of the form
 $$
 \zeta = \zeta_1 + \varepsilon \zeta_2 + \dots,
 $$
 and similar expansions for other variables, where
 \begin{equation}
 \zeta_1 = A(\xi, R, \theta) \phi(z, \theta), \label{11}
 \end{equation}
 with the following set of fast and slow variables:
 \begin{eqnarray}
 \xi = r k(\theta) - s t, \quad R = \varepsilon r k(\theta), \quad \theta = \theta,
 \end{eqnarray}
 where we define $s$ to be the wave speed in the absence of a  shear flow (with $k(\theta) = 1$), while for a given shear flow the function $k(\theta)$ describes the distortion of the wavefront and is to be determined.  In this description, when a shear flow is present, the wave speed in the direction $\theta$ is not equal to $s$, but to $\frac{s}{k(\theta)}$.  The choice of the fast and slow variables is similar to that in the derivation of the cKdV-type equation for the surface waves \citep{Johnson90}, with the formal range of asymptotic validity of the model being defined by the conditions $\xi\sim R\sim O(1)$.  To leading order, the wavefront at any fixed moment of time $t$ is described by the equation
$$rk(\theta)=\text{constant},$$
and we consider  outward propagating ring waves, requiring that the function $k(\theta)$ is strictly positive.  By writing (\ref{11}) we anticipate that the solution of the linearised problem allows for a modal decomposition, similar to the well-known result in the Cartesian geometry, but expect that it has a more complicated structure for the ring waves on a shear flow because of the loss of the radial symmetry in the problem formulation (the shear flow is horizontal). 
 
 To leading order, assuming that perturbations of the basic state are caused only by the propagating wave, we obtain 
 \begin{eqnarray}
&& u_1 = - A \phi u_{0z} \cos \theta - \frac{k F}{k^2 + k^{'2}} A\phi_z, \label{O1_1} \\
&& v_1 = A \phi u_{0z} \sin \theta -  \frac{k' F}{k^2 + k^{'2}} A\phi_z,  \label{O1_2}\\
&& w_1 =  A_{\xi} F \phi,  \label{O1_3} \\
&& p_1 = \frac{\rho_0}{k^2 + k^{'2}} A F^2 \phi_z,  \label{O1_4}  \\
&& \rho_1 = - \rho_{0z} A \phi,  \label{O1_5}  \\
&&\eta_1 = A \phi \quad \mbox{at} \quad z = 1,  \label{O1_6} 
\end{eqnarray}
where the function $\phi(z, \theta)$ satisfies the following modal equations:
\begin{eqnarray}
&&\left (\frac{\rho_0 F^2}{k^2 + k^{'2}} \phi_z\right )_z -  \rho_{0z} \phi = 0, \label{m1} \\
&&\frac{F^2}{k^2 + k^{'2}} \phi_z -  \phi = 0 \quad \mbox{at} \quad z=1, \label{m2} \\
&&\phi = 0 \quad \mbox{at} \quad z=0, \label{m3}
\end{eqnarray}
and
$$
F = -s + (u_0 - c) (k \cos \theta - k' \sin \theta),\label{F}
$$
 where we now have fixed the speed of the moving coordinate frame $c$ to be equal to the speed of the shear flow at the bottom, $c =u_0(0)$ (then, $F = -s \ne 0$ at $z = 0$, and the condition $F \phi = 0$ at $z=0$ implies (\ref{m3})). Of course, the physics does not depend on the choice of $c$ (see a discussion in \cite{Johnson90}), but our derivation shows that the mathematical formulation simplifies if we choose $c = u_0(0)$.
For a given density stratification the values of the wave speed $s$, found assuming the absence of the shear flow, and the pair of functions $\phi(z, \theta)$ and $k(\theta)$, found for a given shear flow, constitute solution of the modal equations (\ref{m1}) - (\ref{m3}). 
Unlike the surface wave problem considered by \cite{Johnson90}, the exact form of equations for the wave speed $s$ and the function $k(\theta)$ depend on stratification. An example for the case of a two-layer fluid with the piecewise - constant current is discussed in section 3. 

At $O(\varepsilon)$ we obtain the following set of equations:
\begin{eqnarray}
&&\rho_0 \left ( F u_{2\xi} + u_{0z} w_2 \cos \theta \right ) + \rho_1 \left ( F u_{1\xi} + u_{0z} w_1 \cos \theta \right )
 + \rho_0 \left [ (F+s) u_{1R} \qquad\qquad\qquad \right .  \nonumber \\
&&  - (u_0 - c) (u_{1\theta} - v_1) \frac kR \sin \theta + (k u_1 + k' v_1) u_{1\xi} + \left.u_{1z} w_1 \right ] + k (p_{2\xi} + p_{1R}) = 0 \label{O2_1}, \\
&& \rho_0 \left ( F v_{2\xi} - u_{0z} w_2 \sin \theta \right ) + \rho_1 \left ( F v_{1\xi} - u_{0z} w_1 \sin \theta \right )
 + \rho_0 \left [ (F+s) v_{1R}  \right .  \nonumber \\
&&  - (u_0 - c) (v_{1\theta} + u_1) \frac {k \sin \theta}{R} + (k u_1 + k' v_1) v_{1\xi} + \left . v_{1z} w_1 \right ] +\frac {kp_{1\theta}} R + k' (p_{2\xi} + p_{1R}) = 0 \label{O2_2}, \ \   \\
&& p_{2z} + \rho_2 + \rho_0 F w_{1\xi} = 0, \label{O2_3}\\
&& F \rho_{2\xi} + \rho_{0z} w_2 + (F+s) \rho_{1R} - (u_0 - c) \frac k R \rho_{1 \theta} \sin \theta + (k u_1 + k' v_1) \rho_{1\xi} + w_1 \rho_{1z} = 0, \label{O2_4} \\
&& k u_{2\xi} + k' v_{2\xi} + w_{2z} + k u_{1R}  + \frac kR (v_{1\theta} + u_1) + k' v_{1R} = 0, \label{O2_5}\\
&& F \zeta_{2\xi} - w_2 + (F+s) \zeta_{1R} - (u_0 - c) \frac kR \zeta_{1\theta}  \sin \theta + (k u_1 + k' v_1) \zeta_{1\xi} + w_1 \zeta_{1z} = 0, \label{O2_6} 
 \end{eqnarray}
 and boundary conditions:
 \begin{eqnarray}
 && w_2 = 0 \quad \mbox{at}\quad  z=0, \label{O2_7}\\
&& p_2 = \rho_0 \eta_2 + \frac 12 \rho_{0z} \eta_1^2 - \eta_1 p_{1z}\quad \mbox{at}\quad  z=1, \label{O2_8}\\
&& w_2 = F \eta_{2\xi} + (F+s) \eta_{1R} - (u_0-c) \frac kR  \eta_{1\theta} \sin \theta + (k u_1 + k' v_1) \eta_{1\xi}\nonumber \\
&& \qquad \qquad \quad + F_z \eta_1 \eta_{1\xi} - \eta_1 w_{1z} \quad \mbox{at}\quad  z=1, \label{O2_9}\\
&& \zeta_2 = \eta_2 - \eta_1 \zeta_{1z} \quad \mbox{at}\quad  z=1. \label{O2_10}
 \end{eqnarray}
 Substituting the leading order solution (\ref{11}), (\ref{O1_1})-(\ref{O1_3}) into the equation (\ref{O2_6}) we obtain
 \begin{equation}
 w_2 = F \zeta_{2\xi} + (F+s) A_R \phi - (u_0-c) \frac kR  \sin \theta\  (A \phi)_{\theta} - F_z \phi^2 A A_{\xi}.
 \label{O2_11}
 \end{equation}
 Next, we find $u_{2\xi}$ from (\ref{O2_1}) and $v_{2\xi}$ from (\ref{O2_2}) and substitute them into (\ref{O2_5}), obtaining the following equation:
 \begin{eqnarray}
&& -(k^2+k'^2) p_{2\xi} + \rho_0 (F w_{2z} - F_z w_2) = \rho_0 \left \{ -\frac kR F (v_{1\theta} + u_1) - (u_0-c) \frac kR \sin \theta [k (u_{1\theta} - v_1) \right . \nonumber \\
&&\left .   + k' (v_{1\theta} + u_1)] + s(k u_{1R} + k' v_{1R}) + \frac 12 [(k u_1 + k' v_1)^2]_{\xi} + (k u_1 + k' v_1)_z w_1 \right \} \nonumber  \\
&&- \rho_1 (F w_{1z} - F_z w_1)  + (k^2 + k^{'2}) p_{1R} + \frac{k k'}{R} p_{1\theta}.\quad  \label{O2_12}
 \end{eqnarray}
 On the other hand, finding $\rho_2$ from (\ref{O2_3}) and substituting it into (\ref{O2_4}) we get
 \begin{eqnarray}
& F p_{2z\xi} - \rho_{0z} w_2 =&\nonumber \\
&\quad  - \rho_0 F^2 w_{1\xi\xi} &+ (F+s) \rho_{1R} - (u_0 - c) \frac kR \rho_{1\theta} \sin \theta + (k u_1 + k' v_1) \rho_{1\xi}  + \rho_{1z} w_1. \quad  \label{O2_14}
 \end{eqnarray}
 Equating the expressions for $p_{2z\xi}$ from the equations (\ref{O2_12}) and (\ref{O2_14}), using (\ref{O2_11}) to exclude $w_2$, substituting the leading order solution (\ref{11}), (\ref{O1_1}) - (\ref{O1_6}) and using the modal equation (\ref{m1}), we obtain the equation for $\zeta_2$ in the form
 \begin{equation}
 \left ( \frac{\rho_0 F^2}{k^2 + k'^2} \zeta_{2\xi z} \right )_z - \rho_{0z} \zeta_{2\xi} = M_2, \label{NE}
 \end{equation}
 where 
 \begin{eqnarray*}
 &&- (k^2 + k'^2) M_2 =  2 s(\rho_0 F \phi_z)_z A_R +   (k^2 + k'^2) \rho_0 F^2  \phi A_{\xi \xi \xi}
+  [- (3 \rho_0 F^2 \phi_z^2)_z + 2 \rho_0 F^2 \phi_z \phi_{zz} \\
&&+ (2 \rho_0 F^2 \phi \phi_{zz})_z] A A_{\xi}  - \left\{\rho_0 \left \{ k(k+k'')\left[ \frac{F^2}{k^2+k'^2}-\left(\frac{2k'F}{k^2+k'^2}+(u_0-c)\sin\theta\right)^2\right]\phi_z\right.\right.\\
&&\left. \left. +2kF \left( \frac{k'F}{k^2+k'^2}+(u_0-c)\sin\theta\right)\phi_{z\theta} \right \} \right\}_z\frac{A}{R} - \left \{ 2 \rho_0 k F  \left [ \frac{k' F}{k^2 + k'^2} + (u_0-c) \sin \theta \right ] \phi_z \right \}_z \frac{A_\theta}{R}.
 \end{eqnarray*}
 Next, substituting (\ref{O2_11}) into the boundary condition (\ref{O2_7}) and recalling that $\phi |_{z=0} = 0$, we obtain $ F \zeta_{2 \xi} = 0 \quad \mbox{at} \quad z=0$, implying that
 \begin{equation}
 \zeta_{2 \xi} = 0 \quad \mbox{at} \quad z=0, \label{NBC1}
\end{equation}
since $F = -s$ at $z = 0$ by our choice of the constant $c$.
The boundary condition (\ref{O2_10}) implies
\begin{equation}
\eta_2 = \zeta_2 + A^2 \phi \phi_z \quad \mbox{at} \quad z=1. \label{O2_16}
\end{equation}
 Substituting (\ref{O2_16}) into (\ref{O2_8}) we get
 \begin{equation}
 p_2 = \rho_0 (\zeta_2 + A^2 \phi \phi_z) -\frac{A^2 \phi}{k^2+k'^2}(\rho_{0z} F^2 \phi_z+2 \rho_0 FF_z \phi_z+\rho_0 F^2 \phi_{zz})+ \frac 12 \rho_{0z} A^2 \phi^2 \quad \mbox{at} \quad z=1. \label{O2_18}
 \end{equation}
 Substituting the expression (\ref{O2_11}) into (\ref{O2_9}), we obtain 
$$ F\zeta_{2\xi}=F\eta_{2\xi}-2FAA_\xi \phi\phi_z,$$
which coincides with the product of  the partial derivative of $\eta_2$ in (\ref{O2_16}) with respect to $\xi$ with $F$. Therefore, the boundary condition 
(\ref{O2_9}) is the differential consequence of (\ref{O2_16}). 
 
 Differentiating (\ref{O2_18}) with respect to $\xi$, using (\ref{O2_12}) to eliminate $p_{2\xi}$, and using (\ref{O2_9}) to exclude $w_2$ (all at $z=1$), we obtain
 \begin{equation}
 \rho_0 \left [ \frac{F^2}{k^2 + k'^2} \zeta_{2\xi z} - \zeta_{2\xi} \right ] = N_2 \quad \mbox{at} \quad z=1, \label{NBC2}
 \end{equation}
 where
 \begin{eqnarray*}
&&- (k^2 + k'^2) N_2 = 2 s \rho_0 F \phi_z A_R + [-3 \rho_0 F^2 \phi_z^2 + 2 \rho_0 F^2 \phi \phi_{zz}] A A_{\xi} \\
&& - \rho_0 \left \{ k(k+k'')\left[ \frac{F^2}{k^2+k'^2}-\left(\frac{2k'F}{k^2+k'^2}+(u_0-c)\sin\theta\right)^2\right]\phi_z\right.\\
&&\left. +2kF \left( \frac{k'F}{k^2+k'^2}+(u_0-c)\sin\theta\right)\phi_{z\theta} \right \}  \frac{A}{R} - 2 \rho_0 k F  \left [ \frac{k' F}{k^2+k'^2} + (u_0-c) \sin \theta \right ] \phi_z \frac{A_\theta}{R}.
\end{eqnarray*}
Thus, we obtained the non-homogeneous equation (\ref{NE}) for the function $\zeta_{2\xi}$ with the boundary conditions (\ref{NBC1}), (\ref{NBC2}). The compatibility condition
$$
\int_0^1 M_2 \phi ~ \mathrm{d}z - [N_2 \phi]_{z=1} = 0
$$
yields the 2+1-dimensional evolution equation for the slowly varying amplitude of the ring wave in the form
\begin{equation}
\mu_1 A_R + \mu_2 A A_{\xi} + \mu_3 A_{\xi \xi \xi} + \mu_4 \frac{A}{R} + \mu_5 \frac{A_\theta}{R} = 0, \label{cKdV}
\end{equation}
where the coefficients are given in terms of the solutions of the modal equations (\ref{m1}) - (\ref{m3}) by the following formulae:
\begin{eqnarray}
&&\mu_1 = 2 s \int_0^1 \rho_0 F \phi_z^2 ~ \mathrm{d}z,  \label{c1}\\
&&\mu_2 = - 3 \int_0^1 \rho_0 F^2 \phi_z^3 ~ \mathrm{d}z,   \label{c2}\\
&&\mu_3 = - (k^2 + k'^2) \int_0^1  \rho_0 F^2 \phi^2 ~ \mathrm{d}z,   \label{c3}\\
&&\mu_4 = - \int_0^1 \left \{  \frac{\rho_0 \phi_z^2 k (k+k'')}{(k^2+k'^2)^2} \left ( (k^2-3k'^2) F^2 -
{4 k' (k^2 + k'^2) (u_0-c)\sin \theta} F  \right .  \right . \nonumber \\
&&\left . \left .  - \sin^2 \theta (u_0-c)^2(k^2 + k'^2)^2 \right ) 
  +  \frac{2 \rho_0 k}{k^2 + k'^2} F \phi_z \phi_{z\theta} (k' F + (k^2 + k'^2) (u_0-c) \sin \theta ) \right \} ~ \mathrm{d}z,\qquad   \label{c4}\\
&& \mu_5 = - \frac{2k}{k^2 + k'^2} \int_0^1 \rho_0 F \phi_z^2 [k' F + (u_0-c) (k^2+k'^2) \sin \theta ] ~ \mathrm{d}z.  \label{c5}
\end{eqnarray}

Let us now consider a reduction to the case of surface ring waves in a homogeneous fluid, studied by \cite{Johnson80,Johnson90}. Here, $\rho_0$ is a constant and we normalise $\phi$ by setting $\phi=1$ at $z=1$. The wave speed $s$, in the absence of a shear flow, as well as  the function $k(\theta)$ and the modal function $\phi$,  for a given shear flow,  can be easily found from the modal equations (\ref{m1}) - (\ref{m3}).  Indeed, the modal function $\phi$ is given by
\begin{equation}
\phi =  (k^2 + k'^2) \int_0^z \frac{1}{F^2} \,\mathrm{d}z. \label{mf} 
\end{equation}
Assuming first that  there is no shear flow, we set $k=1$, $u_0(z)=c=0$, and the normalisation condition $\phi=1$ at $z=1$ then implies
$$\int _0^1 \frac{1}{s^2} \,\mathrm{d} z=1\quad  \Longrightarrow \quad s^2=1.$$
Thus, we recover that the wave speed of the outward propagating surface ring wave in the absence of a shear flow is equal to $1$. Next, let us assume that there is a shear flow, then $F=-1+(u_0-c)(k\cos\theta-k'\sin\theta)$. 
Requiring again that $\phi=1$ at $z=1$ we obtain the equation
\begin{equation}
(k^2+k'^2)\int_0^1\frac {1}{F^2}\,\mathrm{d}  z=1,
\label{gbcon}
\end{equation}
which, of course, coincides with the generalised Burns condition \citep{Johnson90}.

The generalised Burns condition (\ref{gbcon}) constitutes a nonlinear first - order differential equation for the function $k(\theta)$.  
The function relevant to the ring wave is provided by the singular solution  (the envelope of the general solution) of this equation  \citep{Johnson90,Johnson_book}:
\begin{eqnarray}
&&k(\theta)=a \cos(\theta)+b(a)\sin(\theta), \label{gs} \\
&&0=\cos\theta+b'(a)\sin\theta,\\
&&[a^2+b^2(a)]\int_0^1\frac{\mathrm{d}z}{[1-(u_0 (z)-c) a]^2}=1.
\end{eqnarray} 
Indeed, the general solution (\ref{gs}) can not describe an outward propagating ring wave since $k(0)$ can not be strictly positive (see \cite{Johnson90, Johnson_book}).

Differentiating the generalised Burns condition, we obtain:
$$2 (k+k'')  \int_0^1\frac{k'F+(u_0-c)(k^2+k'^2)\sin\theta}{F^3} \,\mathrm{d}  z=0.$$
Since  $k+k''\ne 0$ on the singular solution, we have
$$\mu_5=- 2\rho_0 k (k^2+k'^2)  \int_0^1\frac{k'F+(u_0-c)(k^2+k'^2)\sin\theta}{F^3} \,\mathrm{d}  z=0.$$

Thus, the amplitude equation reduces to the form of the 1+1 - dimensional cKdV-type equation (i.e. $\mu_5 = 0$) for {\it any shear flow}, and not just for stationary and constant shear flows, as previously thought \citep{Johnson90,Johnson_book}. 

Substituting the modal function (\ref{mf}) into the expressions for the remaining coefficients (\ref{c1}) - (\ref{c4}), the amplitude equation can be written as
\begin{equation}
\tilde \mu_1 A_R+\tilde \mu_2 AA_\xi+\tilde \mu_3 A_{\xi\xi\xi} +\frac{\tilde \mu_4}{R} A=0,
\end{equation}
and the expressions for the coefficients can be brought to the previously known form \citep{Johnson90}:
\begin{eqnarray*}
&&\tilde \mu_1= 2(k^2+k'^2) I_3, \\
&&\tilde \mu_2= -3(k^2+k'^2)^2I_4,\\
&&\tilde \mu_3= -(k^2+k'^2)^2 \int_{0}^1 \int_{z}^1 \int_{0}^{z_1}\frac{F^2(z_1,\theta)}{F^2(z,\theta)F^2(z_2,\theta)} \,\mathrm{d}  z_2 \,\mathrm{d}  z_1 \,\mathrm{d}  z,\\
&&\tilde \mu_4= \frac{-k(k+k'')}{k\cos \theta-k'\sin\theta}((k\cos\theta+3k'\sin\theta)I_2+4k'\sin\theta I_3)\\
&& \quad -\frac{3k(k+k'')(k^2+k'^2)\sin^2\theta}{(k\cos\theta-k'\sin\theta)^2}(I_2+2I_3+I_4),\\
&& \mbox{ where }\quad I_n=\int_0^1\frac{\,\mathrm{d}  z}{F^n},\quad \tilde \mu_i=\frac{\mu_i}{k^2+k'^2},~~i=\overline{1,4}.
\end{eqnarray*}
We note that unlike the formula for $\tilde \mu_3$ above, our representation of the coefficients by the formulae (\ref{c1}) - (\ref{c4}) does not involve multiple integrals.

Let us now consider another reduction. If there is no shear flow, i.e. $u_0(z)=0$, the ring waves become concentric waves. Here, 
$F=-s$, $k(\theta)=1$, and $k'(\theta)=0$. Then, the  derived equation (\ref{cKdV}) again reduces  to the form of the 1+1-dimensional cKdV-type equation
\begin{equation}
\tilde \mu_1 A_R+\tilde\mu_2 AA_\xi+\tilde\mu_3 A_{\xi\xi\xi} +\frac{\tilde\mu_4}{R} A=0,
\end{equation}
and expressions for the coefficients are given by
\begin{eqnarray*}
&&\tilde\mu_1=2\int_{0}^1 \rho_0\phi_z^2 \,\mathrm{d} z,\quad
\tilde\mu_2=3\int_{0}^1 \rho_0\phi_z^3 \,\mathrm{d} z,\\
&&\tilde\mu_3=\int_{0}^1\rho_0\phi^2 \,\mathrm{d}  z,\quad
\tilde\mu_4=\int_{0}^1  \rho_0 \phi_z^2  \,\mathrm{d} z,\\
&& \mbox{ where } \tilde \mu_i=-\frac{\mu_i}{s^2},~~i=\overline{1,4},
\end{eqnarray*}
where $\phi= \phi(z)$ is the modal function, satisfying the same equations as in Cartesian geometry:
\begin{eqnarray*}
&&(s^2 \rho_0 \phi_z)_z - \rho_{0z} \phi = 0, \\
&&s^2 \phi_z - \phi = 0 \quad \mbox{at} \quad z=1, \\
&&\phi = 0 \quad \mbox{at} \quad z=0.
\end{eqnarray*}
This reduction agrees with  the equation previously derived by Lipovskii \citep{Lipovskii85}.

 
Thus, in both cases the derived amplitude equation (\ref{cKdV}) correctly reduces to the previously derived models. Importantly, in both of these previously studied cases the coefficient $\mu_5 = 0$. However, it is not equal to zero in the general case of the waves in a stratified fluid with a shear flow. Interestingly, Johnson has caught the `ghost' of the additional term in his study of the surface waves on a shear flow. However, as we discussed earlier, the complicated formula for the coefficient $\mu_5$ given by Johnson \citep{Johnson90} will lead to a zero coefficient  for any shear flow.


\section { Two-layer example: dispersion relation and wavefronts}
\subsection{Two-layer Model}
In order to clarify the general theory developed in the previous section and to illustrate the different effect of a shear flow on the wavefronts of surface and internal ring waves, here we discuss a simple piecewise-constant setting,  frequently used in theoretical and laboratory studies of long internal and surface waves  (Figure 3, see, for example,  \cite{Long55, Lee, Miyata85, Weidman, CC, Ramirez, Choi, Grue, Voronovich, BC, Chumakova, Boonkasame, AGK, ASK} and references therein). 
In these theoretical and laboratory studies, the model is often chosen to yield explicit formulae, but is regarded as a reasonable abstraction for a background flow with smooth density and shear profiles across the interface, in the long wave approximation. It is also necessary to note that this background flow is subject to Kelvin-Helmholtz (K-H) instability arising as short waves, which are excluded in the long wave theory due to the large separation of scales  (see \cite{Drazin, Turner, Craik} and references therein).
\begin{figure}
           \centering\includegraphics[width=0.6\columnwidth]{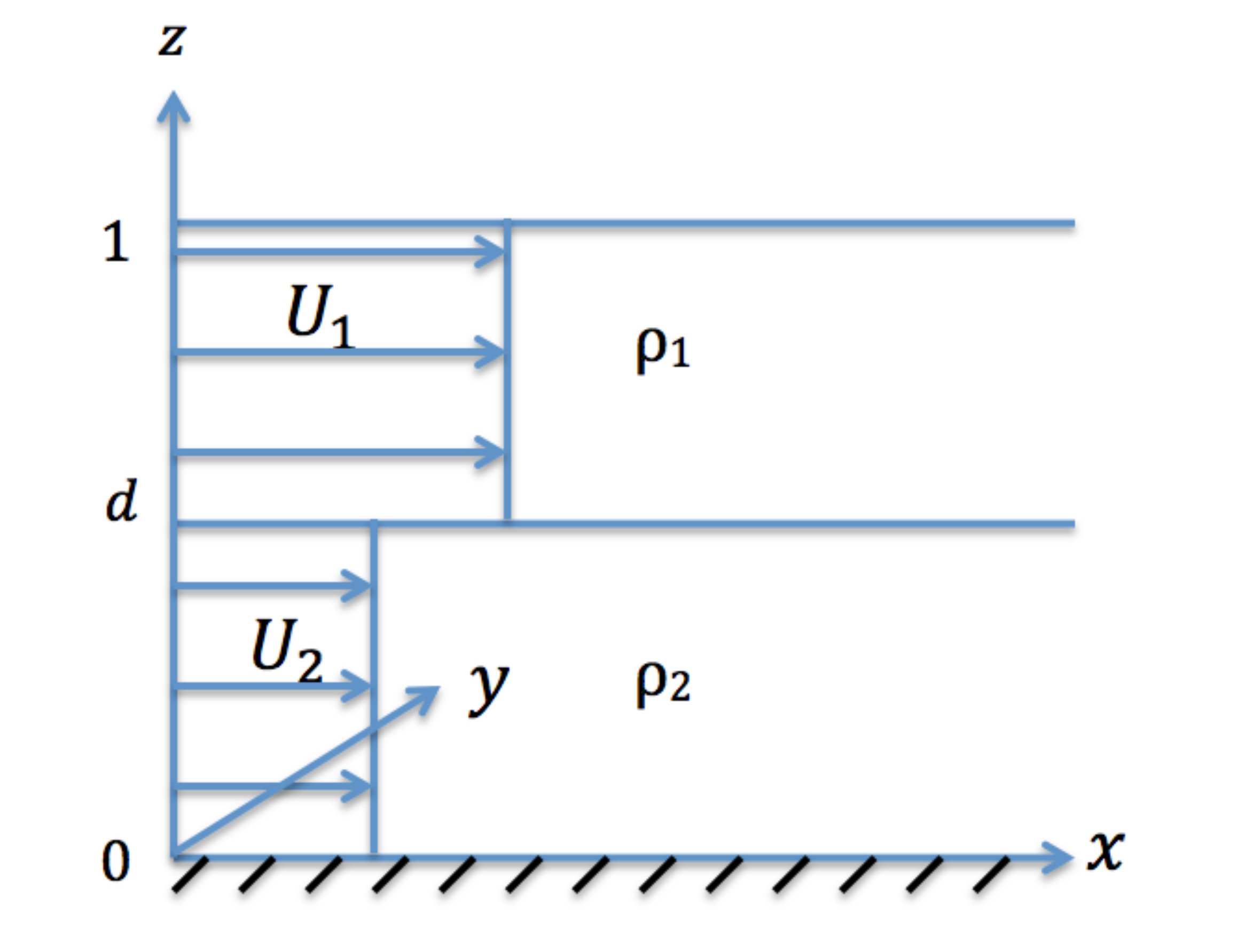} 
		\caption {Two-layer model.}
\end{figure}

Here, in non-dimensional form, both the density of the fluid and the shear flow are piecewise-constant functions ($0 \le z \le 1$):	
\begin{eqnarray*}
&&\rho_0 = \rho_2 H(z) + (\rho_1 - \rho_2) H(z-d), \\
&&u_0 = U_2 H(z) + (U_1 - U_2) H(z-d),
\end{eqnarray*}
where $d$ is the thickness of the lower layer and $H(z)$ is the Heaviside function. 

In the rigid lid approximation, the condition for the stability of the long waves in a fluid of finite depth in nondimensional variables used in our paper is given by 
\begin{equation}
(U_1 - U_2)^2 < \frac{(\rho_2-\rho_1)\big( \rho_1 d + \rho_2 (1-d) \big)}{\rho_1\rho_2}\label{ucl}
\end{equation}
\citep{Ovsyannikov, Ovsyannikov_book}, see also \cite{Bontozoglou91, BM, Lannes} and references therein. The free surface results show that long waves are stable both for small shears, as in the rigid lid case, but also for sufficiently large shears \citep{Ovsyannikov, Ovsyannikov_book}, see also \cite{ BarrosChoi, Lannes}. 

Long interfacial waves are observed both in experiments and in natural environments (for example, see reviews by \cite{Grue, Apel07}), which means that in the situations when they are observed there exist some extra mechanisms which prevent the development of the K-H instability. Among such stabilising mechanisms not present in the simplified model are the continuity of the actual density and shear flow profiles, with a thin intermediate layer in between the two main layers, and surface tension (see \cite{Turner, BM, Lannes} and references therein). The two-layer model is an abstraction of the actual continuous density and shear flow profiles, suitable for  the theoretical study of the {\it long} internal waves.

\subsection{Dispersion relation and approximations}

Solution of the modal equations  (\ref{m1}) - (\ref{m3}) in the two layers is given by the linear functions of $z$ ($\phi_1$ is the modal function in the upper layer and $\phi_2$ is the modal function in the lower layer):
\begin{eqnarray}
\phi_1 = \left ( \frac{F_1^2}{k^2 + k'^2} + z - 1 \right ) {\Lambda}, \quad
\phi_2 =  \left ( \frac{F_1^2}{k^2 + k'^2} + d - 1 \right ) \frac{\Lambda z}{d},\label{phi}
\end{eqnarray}
where $\Lambda$ is a constant and the continuity of $\phi$ is satisfied, while the jump condition at the interface
$$
\frac{\left [ \rho_0 F^2 \phi_z \right ]}{k^2 + k'^2} = [\rho_0] \phi \quad \mbox{at} \quad z=d
$$
provides the `dispersion relation',
\begin{equation}
(\rho_2 - \rho_1) d (1-d) (k^2 + k'^2)^2 - \rho_2 [d F_1^2 + (1-d) F_2^2] (k^2 + k'^2) + \rho_2 F_1^2 F_2^2 = 0,
\label{k}
\end{equation}
with $F_1 = -s + (U_1 - U_2) (k \cos \theta - k' \sin \theta), F_2 = -s$.
This nonlinear first-order differential equation for the function $k(\theta)$ is further generalisation of both Burns and generalised Burns conditions \citep{ Burns53, Johnson90}.  
First, we assume that there is no shear flow and find the wave speed $s$ by letting $U_1=U_2=0$, while $k=1$. The dispersion relation takes the form
$$\rho_2 s^4 -\rho_2 s^2 + (\rho_2-\rho_1)d (1-d)=0.$$
So the wave speed in the absence of the shear flow is given by
$$s^2=\frac{\rho_2\pm \sqrt{\rho_2^2-4\rho_2(\rho_2-\rho_1) d (1-d)}}{2\rho_2}=\frac{1\pm \sqrt{(2d-1)^2+4\rho_1/\rho_2 d(1-d)}}{2},$$
where the upper sign should be chosen for the faster surface mode, and the lower sign for the slower internal mode. For example, if $\rho_1 = 1, \rho_2 = 1.2$ and $d=0.5$, we obtain $s_{sur} \approx 0.98$ and $s_{int} \approx 0.21$.

When the shear flow is present, equation (\ref{k}) constitutes a nonlinear first-order differential equation for the function $k(\theta)$. We have 
\begin{equation}
k^2+k'^2=\frac{\rho_2[dF_1^2+(1-d)F_2^2]\pm\sqrt{\Delta}}{2(\rho_2-\rho_1)(1-d)d},
\label{kk}
\end{equation}
where
\begin{eqnarray*}
\Delta &=& \rho_2^2 [d F_1^2 + (1-d) F_2^2]^2 - 4 \rho_2 (\rho_2 - \rho_1) d (1-d) F_1^2 F_2^2\\
&=&  \rho_2^2 [d F_1^2 - (1-d) F_2^2]^2 + 4 \rho_1 \rho_2 d (1-d) F_1^2 F_2^2 \ge 0,
\end{eqnarray*}
and the upper / lower signs correspond to the internal / surface modes, respectively.

The generalised Burns condition for the surface waves in a homogeneous fluid with this two-layer shear flow takes the form
\begin{equation}
(k^2+k'^2)\left (\frac{1-d}{F_1^2}+\frac{d}{F_2^2}\right )=1~\Leftrightarrow~[d F_1^2 + (1-d) F_2^2] (k^2 + k'^2) = F_1^2 F_2^2,
\label{gBc}
\end{equation}
and can be recovered from our more general equation (\ref{k}) in the limit $\rho_2 - \rho_1 \to 0$.

If the density contrast is small, $\rho_2 - \rho_1 \ll \rho_1, \rho_2$, one can obtain a simplified equation not only for the surface mode (see (\ref{gBc})), but also for the interfacial mode, by replacing the free surface condition ($\ref{m2}$) with the rigid lid approximation, 
$$
\phi = 0 \quad \mbox{at} \quad z=1.
$$
Then, the modal function $\phi$ in the two layers is found to be
$$
\phi_1 = \Lambda (z-1), \quad \phi_2 = \frac{d-1}{d} \Lambda z,
$$
where $\Lambda$ is a constant, and  the jump condition at $z=d$ again provides the 'dispersion relation':
\begin{equation}
(\rho_2 - \rho_1) d (1-d) (k^2 + k'^2) = \rho_1 d F_1^2 + \rho_2 (1-d) F_2^2.
\label{rl_k}
\end{equation}
Remarkably, the required singular solution $k(\theta)$ of the differential equation (\ref{rl_k}) can be found explicitly, in the following form:
\begin{eqnarray}
&&k(\theta)= \mbox{sign} (\cos\theta)\sqrt{\frac{I^2 K^2 - I (U_1-U_2)+1}{1+(1- I (U_1-U_2) )\tan^2\theta}}\left (\frac{\cos\theta}{1- I (U_1-U_2)}+\frac{\sin^2 \theta}{\cos\theta}\right )\nonumber \\
&&\qquad \qquad\qquad \qquad\qquad \qquad\qquad \qquad\qquad \qquad -\frac{I K \cos\theta}{1-I (U_1-U_2)},\label{Kovy}\\
&& \mbox{where} \qquad I = \frac{\rho_1 (U_1-U_2)}{(1-d)(\rho_2-\rho_1)}, \quad K^2=\frac{(\rho_2-\rho_1)d (1-d)}{\rho_1 d + \rho_2 (1-d)}. \nonumber 
\label{approx}
\end{eqnarray}
The solution (\ref{Kovy}) can be verified directly, by substituting it into the equation (\ref{rl_k}).

Let us note that equations asymptotically equivalent to both approximate equations (\ref{gBc}) and (\ref{rl_k}) can be formally obtained from the full dispersion relation (\ref{k}) if we first solve it as a quadratic equation with respect to $k^2 + k'^2$, see (\ref{kk}), 
and then Taylor expand  $\sqrt{\Delta}$ in powers of the small parameter ${(\rho_2 - \rho_1)}/{\rho_2}$:
$$
k^2 + k'^2 = \frac{\rho_2 [d F_1^2 + (1-d) F_2^2]}{2 (\rho_2 - \rho_1) (1-d) d} \left [1 \pm \left (1 + 2 \frac{(\rho_2 - \rho_1) d (1-d) F_1^2 F_2^2}{\rho_2 [d F_1^2 + (1-d) F_2^2]^2} + \dots \right ) \right ].
$$
The approximate equations  (\ref{gBc}) and (\ref{rl_k}) correspond to the lower and the upper signs in the above equation, respectively.


\subsection{Wavefronts}

The general solution of the equation (\ref{k}) can be found in the form similar to the general solution of the generalised Burns condition \citep{Johnson90}, allowing us then to find the necessary singular solution relevant to the ring waves in a stratified fluid in parametric form:
 \begin{equation}
 \left\{
  \begin{array}{cc}
&k(\theta)=a \cos\theta+b(a)\sin\theta,\\
&b'(a)=-1/\tan\theta,\\
&a^2+b^2=\frac{\rho_2 [d(-s+a(U_1-U_2))^2+(1-d)s^2]\pm \sqrt{\Delta}}{2(\rho_2-\rho_1)d(1-d)},
 \end{array}\right.
 \label{ss}
 \end{equation}
 where
\begin{eqnarray*}
\Delta  &=& \rho_2^2\left [d(-s+a(U_1-U_2))^2-(1-d)s^2\right ]^2+4\rho_1\rho_2d(1-d)s^2\left [-s+a(U_1-U_2)\right ]^2.
\end{eqnarray*}
Therefore,
$$
0 \le \Delta \le \rho_2 [d (-s + a (U_1 - U_2))^2 + (1-d) s^2].
$$
The upper sign should be chosen in (\ref{ss})  for the interfacial mode and the lower sign for the surface mode, as previously discussed.


Let us denote
$$a^2+b^2=\frac{\rho_2 [d(-s+a(U_1-U_2))^2+(1-d)s^2] \pm \sqrt{\Delta}}{2(\rho_2-\rho_1)d(1-d)}=Q.$$
Then the condition $b^2=Q-a^2\ge0$ determines  the domain of $a\in[a_{\text{min}},a_{\text{max}}]$. We require $k(\theta)$ to be positive everywhere to describe the outward propagating ring wave.   Therefore, one needs to choose the interval $[a_{\text{min}},a_{\text{max}}]$ containing $a=0$, since $a$ should take both positive and negative values (in particular, $k(\theta)$ should be positive both at $\theta = 0$ and $\theta = \pi$).
Then,
$$2bb'=Q_a-2a,\quad \Rightarrow b'=\frac{Q_a-2a}{2b},\quad \Rightarrow \tan\theta=-\frac{2b}{Q_a-2a}.$$
Therefore, $k(\theta)$ can be written in the form
$$k(\theta)=a\cos\theta+b\sin\theta=\cos\theta(a+b\tan\theta)=\mbox{sign}\left (-2b \frac{\cos\theta}{\tan\theta}\right )\frac{aQ_a-2Q}{\sqrt{(Q_a-2a)^2+4b^2}}.$$
Since $k(\theta) > 0$, we obtain
$$\mbox{sign}(k(\theta))=-\mbox{sign}\left [b(aQ_a-2Q)\frac{\cos\theta}{\tan\theta}\right ]=1.$$
When there is no shear flow, $aQ_a-2Q = - 2Q <0$. Therefore, it is natural to assume that this inequality will continue to hold in the case of the relatively weak shear flow considered here, and check that it is satisfied once the solution is constructed.  Also, in the particular case of the small density contrast,  $\rho_2 - \rho_1 \ll \rho_{1,2}$, and $d \sim 0.5$, this condition can be verified directly using the approximation (\ref{rl_k}) for the interfacial mode, which we do not discuss in detail here. 

Thus, we assume that $aQ_a-2Q<0$ in the interval  $[a_{\text{min}},a_{\text{max}}]$. Then
$$k(a)=-\frac{aQ_a-2Q}{\sqrt{(Q_a-2a)^2+4b^2}},$$
and
$$ \mbox{sign}(b)=\mbox{sign} \left ( \frac{\cos\theta}{\tan\theta}\right )=\mbox{sign} \left ( \frac{\cos^2\theta}{\sin\theta}\right ) = \left\{ 
 \begin{array}{ll}
     1& \mbox{if} \quad \theta \in  (0,\pi),\\
     -1& \mbox{if} \quad \theta  \in (\pi,2\pi).
   \end{array} \right.$$
Therefore, if $\theta\in(0,\pi)$, then
$$b=\sqrt{Q-a^2},\qquad \tan\theta=-\frac{2\sqrt{Q-a^2}}{Q_a-2a},$$ 
and we let
$$\theta=\left\{
\begin{array}{ll}
\arctan (-\frac{2\sqrt{Q-a^2}}{Q_a-2a})&\quad \mbox{if} \quad Q_a-2a<0,\\
\arctan (-\frac{2\sqrt{Q-a^2}}{Q_a-2a})+\pi&\quad \mbox{if} \quad Q_a-2a>0.
\end{array}\right.$$
Similarly, if  $\theta\in(\pi,2\pi)$, then
$$b=-\sqrt{Q-a^2},\qquad \tan\theta=\frac{2\sqrt{Q-a^2}}{Q_a-2a},$$ 
and  we let
$$\theta=\left\{
\begin{array}{ll}
\arctan (\frac{2\sqrt{Q-a^2}}{Q_a-2a})+\pi&\quad \mbox{if} \quad Q_a-2a>0,\\
\arctan (\frac{2\sqrt{Q-a^2}}{Q_a-2a})+2\pi&\quad \mbox{if} \quad Q_a-2a<0.
\end{array}\right.$$
Thus, we obtained the required singular solution analytically,  in parametric form. The functions $k(a), \theta(a)$ and $k(\theta)$ for both surface and interfacial ring waves are shown in Figures 4, 5 and 6, respectively. As before, we let $\rho_1=1$, $\rho_2=1.2$ and $d=0.5$, and consider several values of the strength of the shear flow.

\begin{figure}
\centering
\renewcommand{\thesubfigure}{(\arabic{subfigure})}
\subfigure[Surface mode: $U_1-U_2=0$ (circles), $U_1-U_2=0.2$ (plus signs), $U_1-U_2=0.3$ (stars), and $U_1-U_2=0.4$ (triangles).]{\includegraphics[width=.48\textwidth]{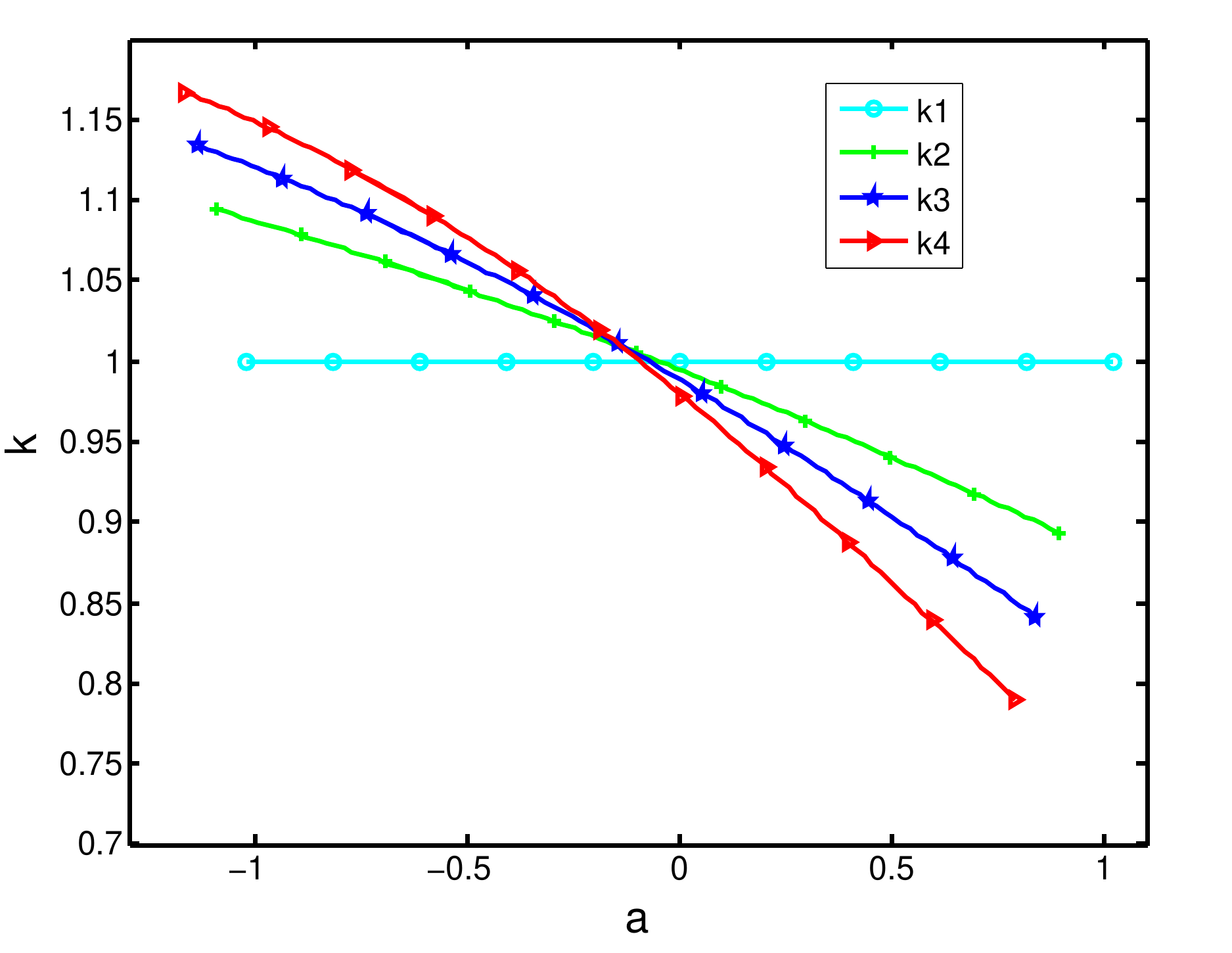}}~~
\subfigure[Interfacial mode: $U_1-U_2=0$ (circles), $U_1-U_2=0.1$ (plus signs), $U_1-U_2=0.15$ (stars), and $U_1-U_2=0.2$ (triangles).]{\includegraphics[width=.48\textwidth]{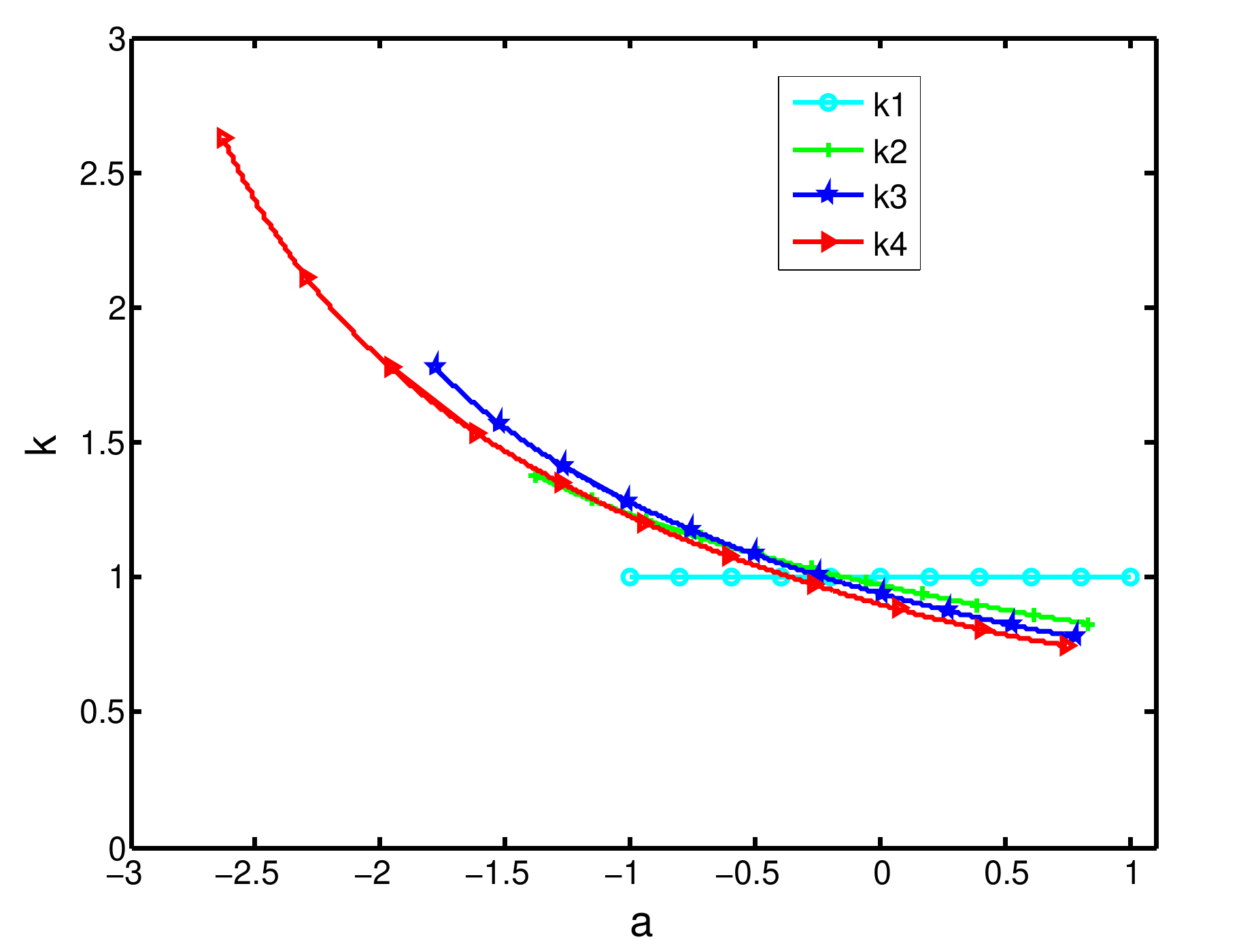}}\\
\caption{Function $k(a)$: surface mode (left) and interfacial mode (right).}
\setcounter{subfigure}{0}
\end{figure}

\begin{figure}
\centering
\renewcommand{\thesubfigure}{(\arabic{subfigure})}
\subfigure[Surface mode: $U_1-U_2=0$ (circles), $U_1-U_2=0.2$ (plus signs), $U_1-U_2=0.3$ (stars), and $U_1-U_2=0.4$ (triangles).]{\includegraphics[width=.48\textwidth]{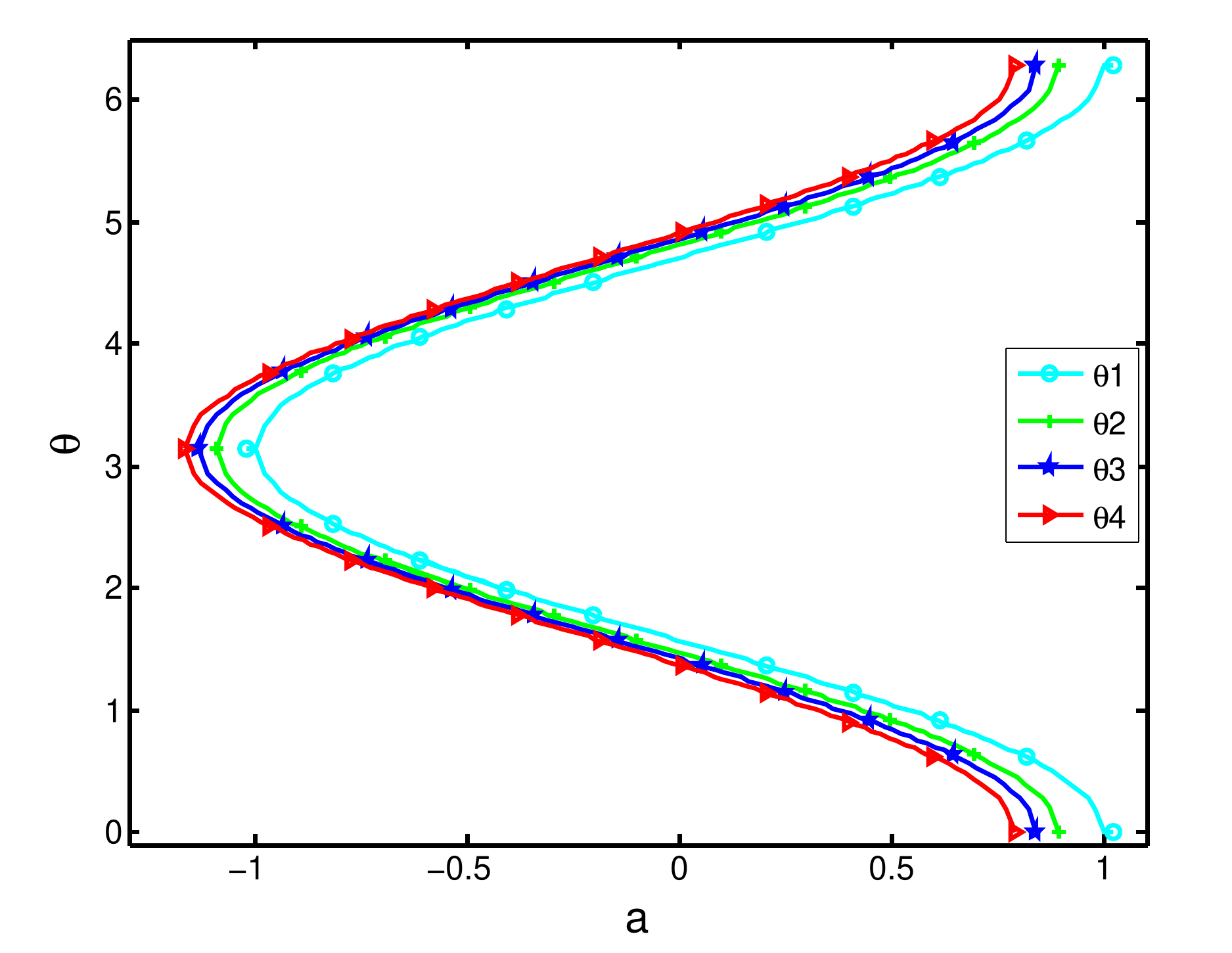}}~~
\subfigure[Interfacial mode: $U_1-U_2=0$ (circles), $U_1-U_2=0.1$ (plus signs), $U_1-U_2=0.15$ (stars), and $U_1-U_2=0.2$ (triangles).]{\includegraphics[width=.48\textwidth]{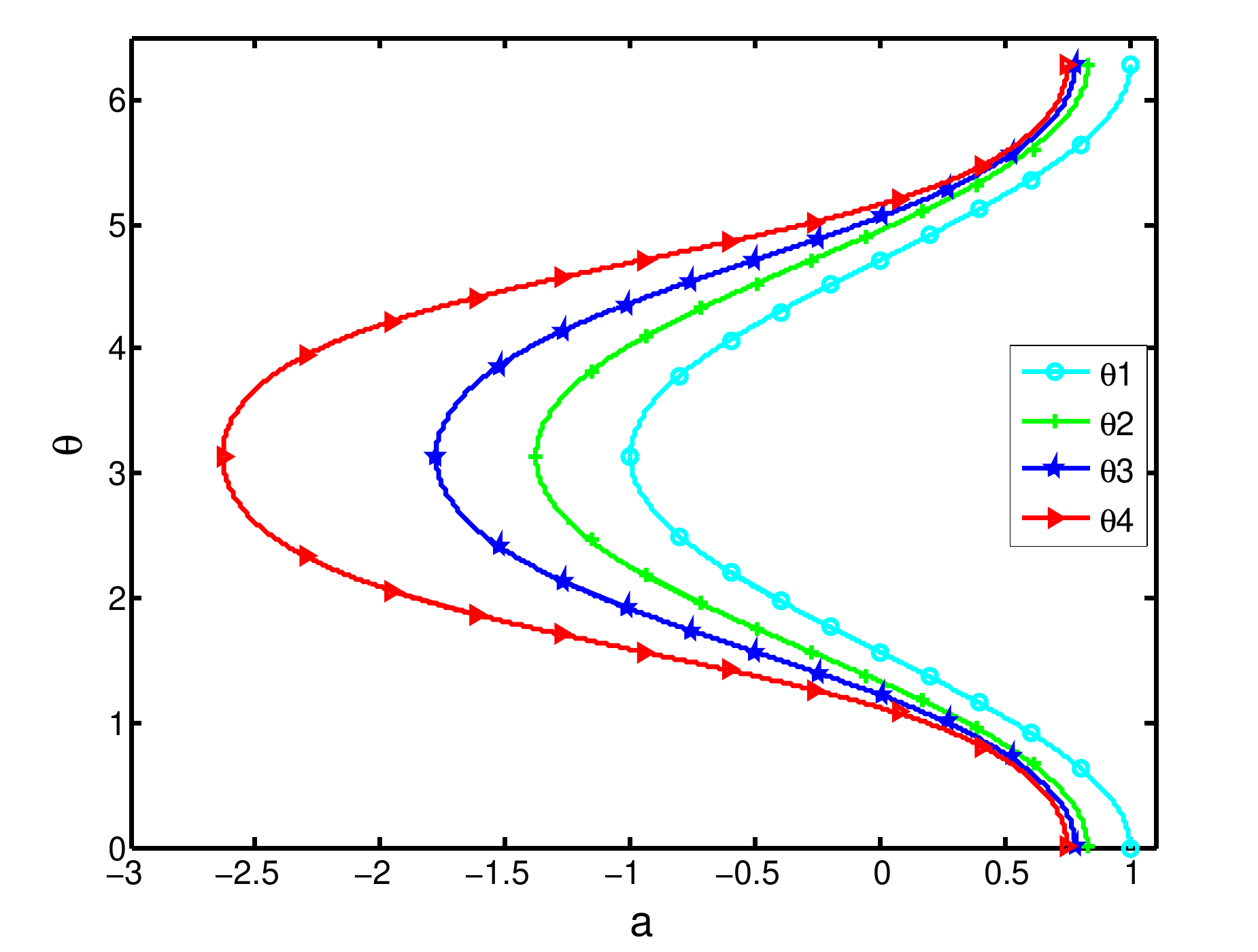}}
\caption{Function $\theta(a)$: surface mode (left) and interfacial mode (right).}
\setcounter{subfigure}{0}
\end{figure}

\begin{figure}
\centering
\renewcommand{\thesubfigure}{(\arabic{subfigure})}
\subfigure[Surface mode: $U_1-U_2=0$ (circles), $U_1-U_2=0.2$ (plus signs), $U_1-U_2=0.3$ (stars), and $U_1-U_2=0.4$ (triangles). ]{\includegraphics[width=.48\textwidth]{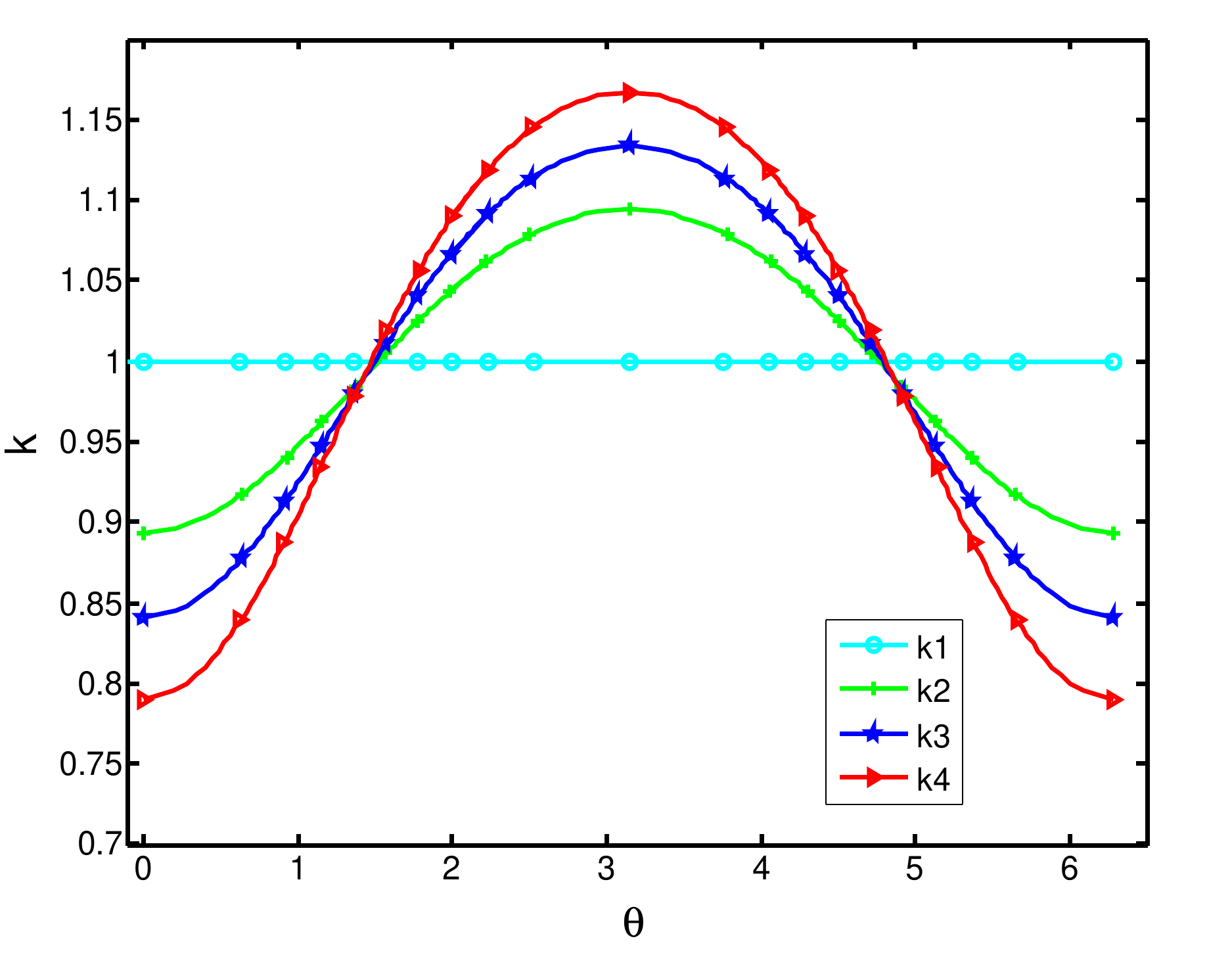}}~~
\subfigure[Interfacial mode: $U_1-U_2=0$ (circles), $U_1-U_2=0.1$ (plus signs), $U_1-U_2=0.15$ (stars), and $U_1-U_2=0.2$ (triangles). ]{\includegraphics[width=.48\textwidth]{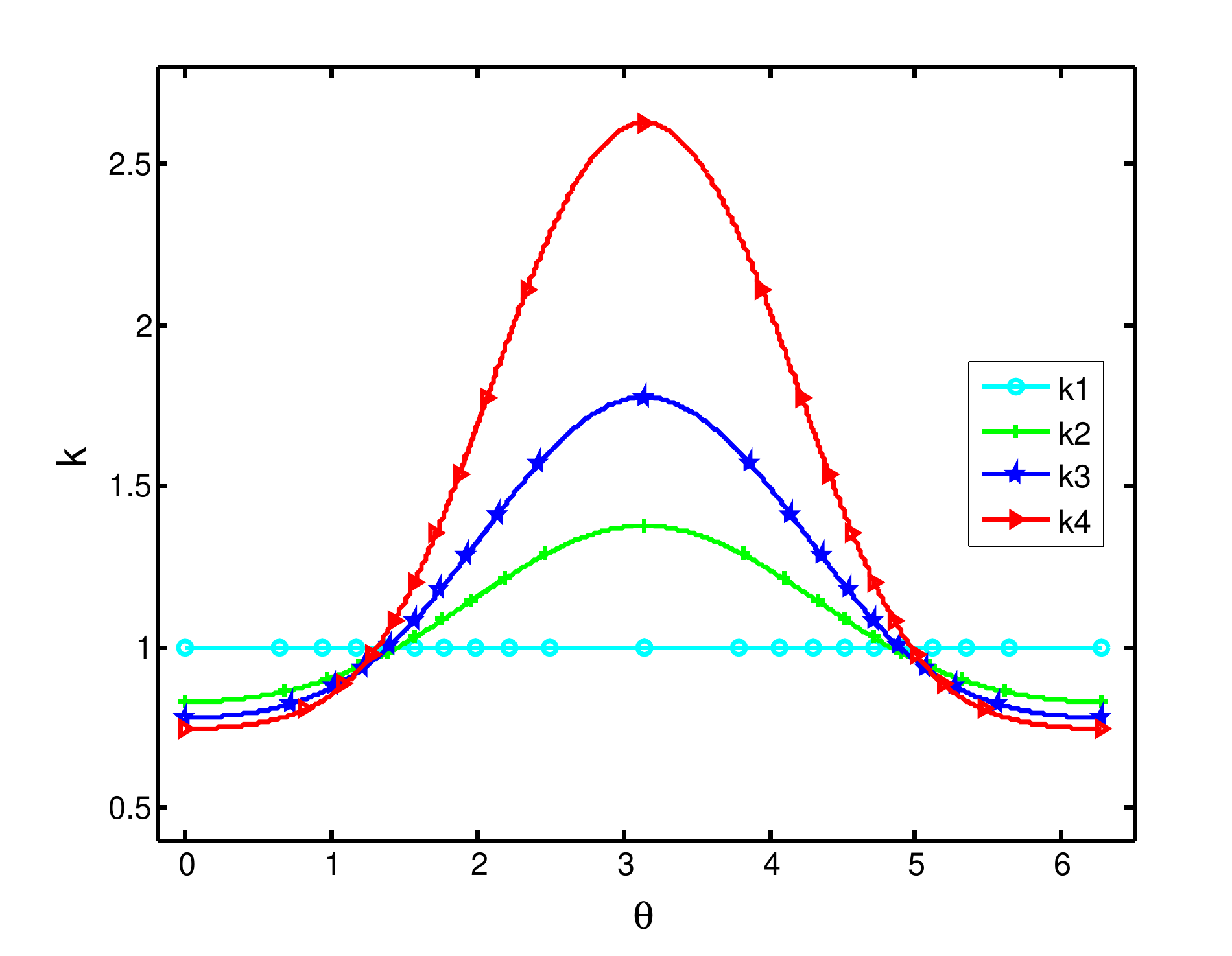}}
\caption{Function $k(\theta)$: surface mode (left) and interfacial mode (right).}
\setcounter{subfigure}{0}
\end{figure}

We also compare the approximate solution (\ref{Kovy}) for the internal waves with the exact solution \ref{ss} in Figure 7,  for the case when $U_1-U_2=0.1$. We can see that the simpler approximate solution is rather close to the exact solution, with the advantage that the function $k$ can be written explicitly  as a function of $\theta$. 

    \begin{figure}
		\centering\includegraphics[width=0.6\columnwidth]{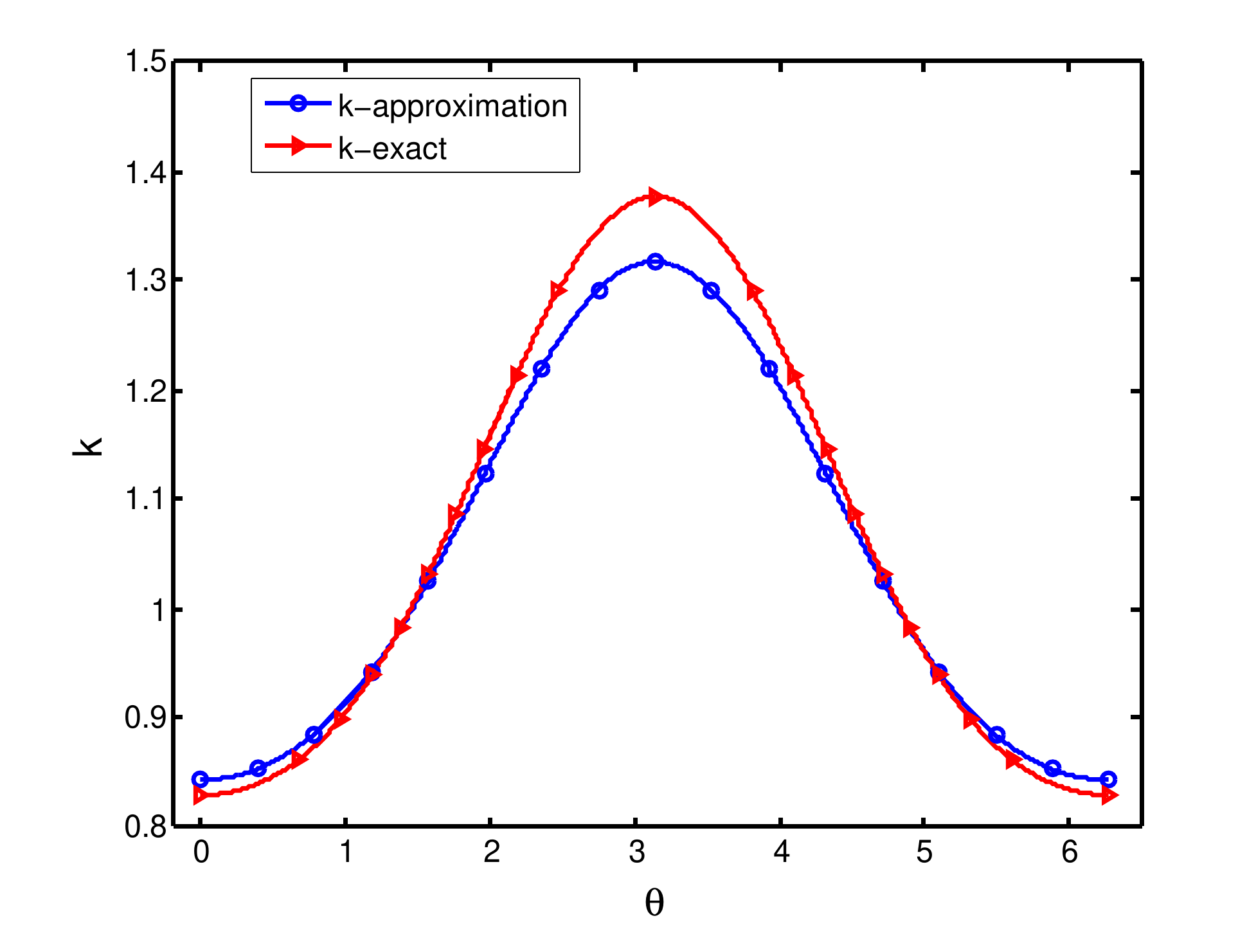} 
		\caption{Function $k(\theta)$: approximate solution (\ref{Kovy}) (circles) and exact solution (\ref{ss})  (triangles) for internal waves; $U_1-U_2=0.1$.}
		\end{figure}

Next, we note that to leading order, the waves propagate at the speed $s/k(\theta)$ in the direction $\theta$. As discussed before, wavefronts are described by the equation $rk(\theta)=$constant.
In Figures 8 and 9 we show the wavefronts for the surface mode and interfacial mode of the equation (\ref{k}), respectively, for $\rho_1 = 1, \rho_2 = 1.2, d = 0.5$ and several values of the strength of the shear flow. 
 
    \begin{figure}
		\centering\includegraphics[width=0.8\columnwidth]{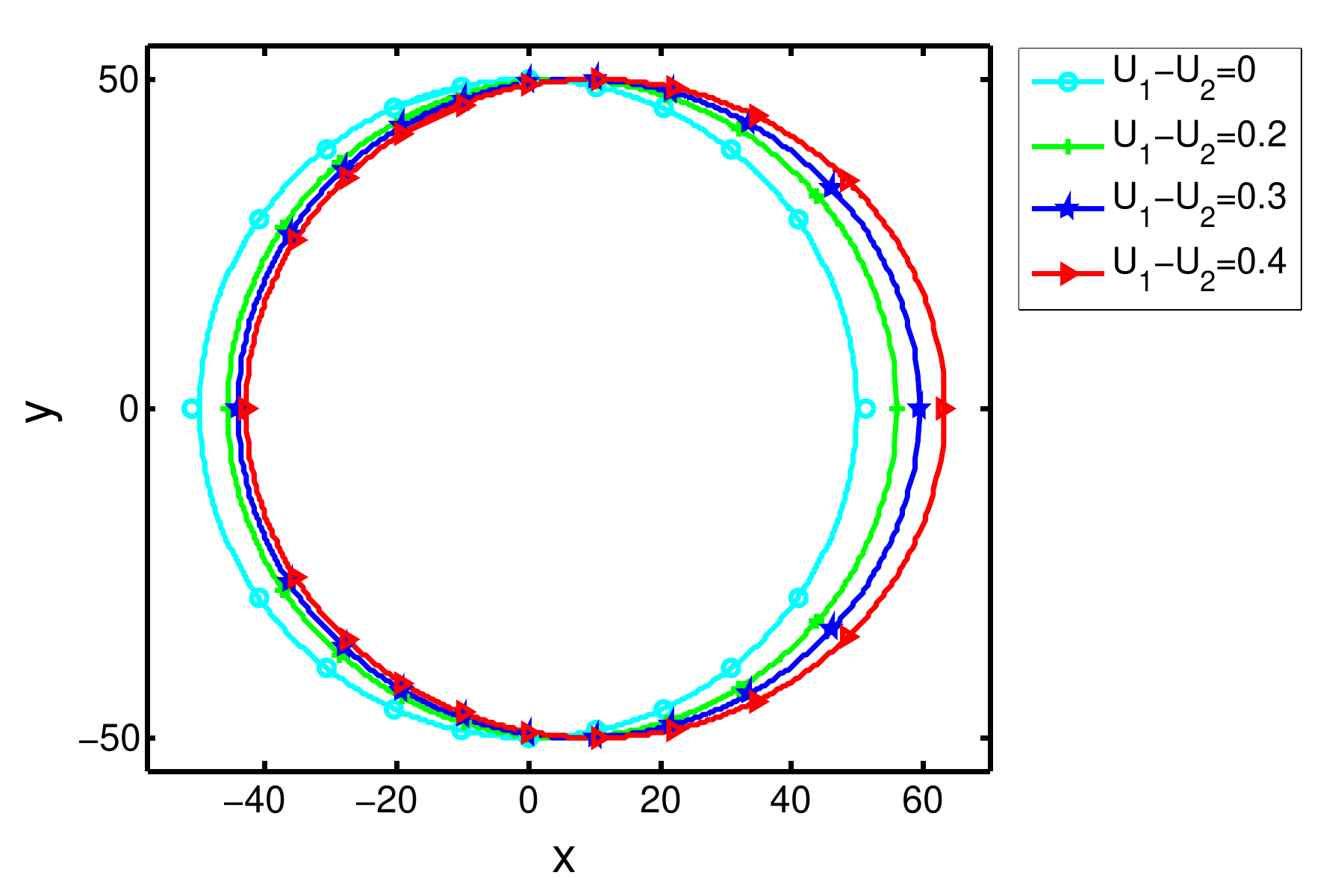} 
		\caption{Wavefronts of surface ring waves described by $k(\theta) r = 50$ ($\varepsilon=0.02$ and $R=1$) for $U_1-U_2=0$ (circles), $U_1-U_2=0.2$ (plus signs), $U_1-U_2=0.3$ (stars), and $U_1-U_2=0.4$ (triangles).}
		\end{figure}
    \begin{figure}
		\centering\includegraphics[width=0.8\columnwidth]{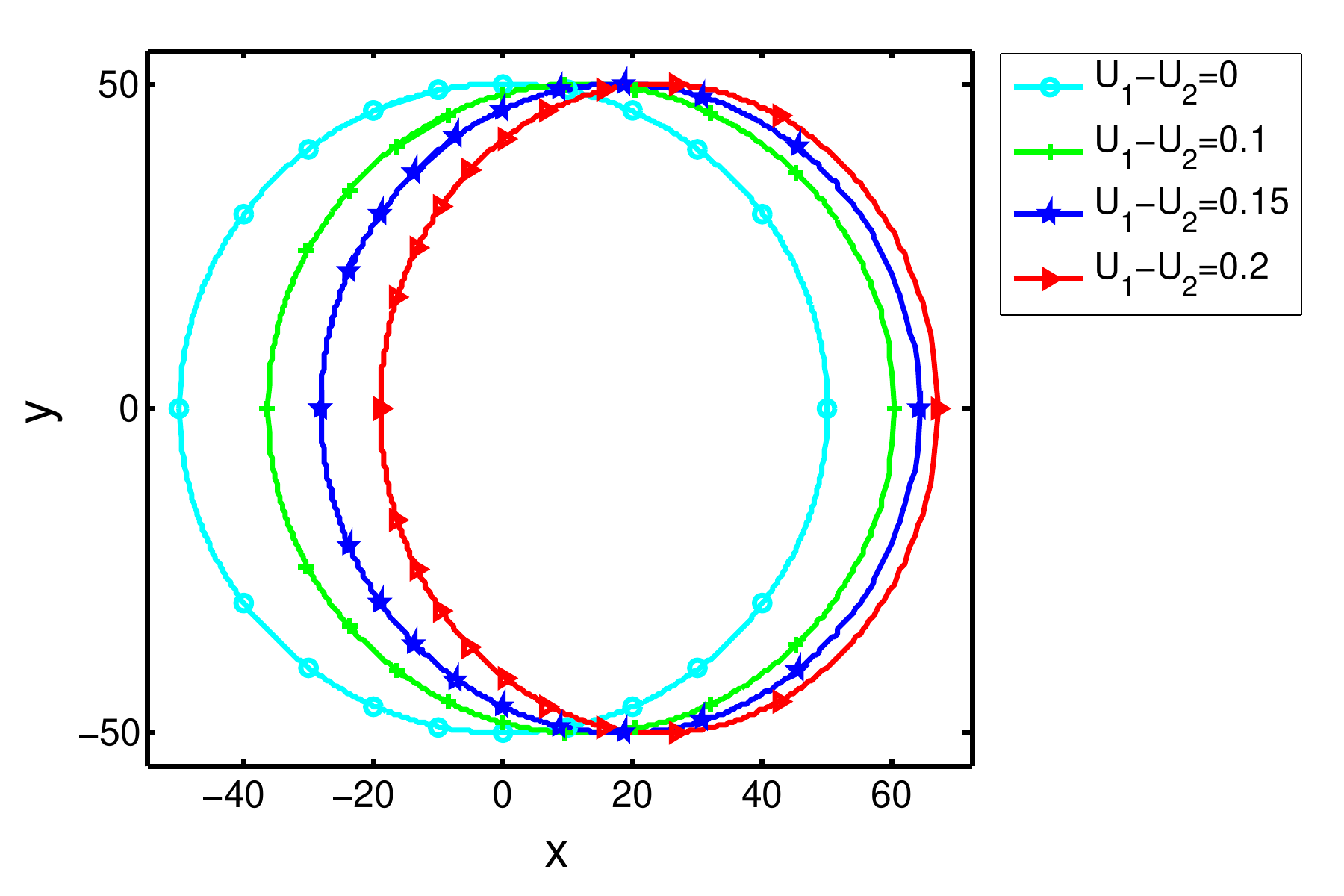} 
		\caption{Wavefronts of interfacial ring waves described by $k(\theta) r = 50$  ($\varepsilon=0.02$ and $R=1$) for $U_1-U_2=0$ (circles), $U_1-U_2=0.1$ (plus signs), $U_1-U_2=0.15$ (stars), and $U_1-U_2=0.2$ (triangles).}
		\end{figure}

We see that the shear flow has very different effect on the surface and internal ring waves: the surface ring waves shown in Figure  8 are elongated in the direction of the shear flow, while the interfacial ring waves shown in Figure  9 are squeezed in the direction of  the flow. We note that, in a different setting, the deformation of surface ship wakes by the sheared current was recently discussed by \cite{Ellingsen}, which is related to the deformation of the surface ring waves. We also note that when the value of $U_1-U_2$ is increased, there is a threshold after which the equation for $k(\theta)$ corresponding to interfacial waves does not have a real-valued solution. This value coincides with the critical value given by (\ref{ucl}).

To understand why this happens, we consider the behaviour of $k(\theta)$ around the angles $\theta = 0, \pi$. 
Locally, in the area around the angles $\theta=0,\pi$, the ring waves can be treated as plane waves over a shear flow (propagating along or opposite the flow). 
The modal equation is given by \citep{Grimshaw01}:
\begin{eqnarray*}
(\rho_0(C-u_0)^2\phi_z)_z-\rho_{0z}\phi&=&0,\\
(C-u_0)^2\phi_z-\phi&=&0,\quad \mbox{at}~ z=1,\\
\phi&=&0,\quad \mbox{at}~z=0.
\end{eqnarray*}
In the two-layer case, the dispersion relation takes the form
$$d\rho_1(C-U_1)^2-\rho_2(C-U_2)^2(d-1+(C-U_1)^2)=d(\rho_1-\rho_2)(d-1+(C-U_1)^2).$$
This dispersion relation also follows from the results obtained by \cite{Ovsyannikov} for the free surface  two-layer shallow water model. 
Substituting the coefficients $\rho_1 = 1,~ \rho_2 = 1.2, ~d = 0.5$ into the general solution of this quartic equation, one obtains the following formula for the wave speed $C$:
\begin{equation}
C=\frac{U_1+U_2}{2}\pm \frac 16 \sqrt{9(U_1-U_2)^2\pm3\sqrt{72(U_1-U_2)^2+30}+18}.
\label{C}
\end{equation}
Letting the Cartesian coordinate frame move at the speed $U_2$, we now plot the four solutions for the wave speed $C$  in Figure 10. Here, the top and bottom curves show the speeds of surface waves, propagating along and opposite the shear flow, while the curves in between show the speeds of the slower moving internal waves. Both surface and internal waves can propagate when there is no shear flow ($U_1 = 0$.) When the strength of the shear flow is increased ($U_1 > 0$), the difference between the speeds of the surface waves along and opposite the flow increases, while a similar difference for the internal waves decreases. This indicates that the wavefronts of the surface waves become elongated in the direction of the flow, while the wavefronts of the internal waves are indeed squeezed in this direction. The graph also shows the onset of the K-H instability for the long waves at $U_1 \approx 0.5$ and stabilisation for the values of the shear flow exceeding $U_1 \approx 2$, in agreement with the results of \cite{Ovsyannikov}. We note that the stabilisation persists within the scope of the full equations of motion \citep{Ovsyannikov_book, Lannes}. 

    \begin{figure}
		\centering\includegraphics[width=0.7\columnwidth]{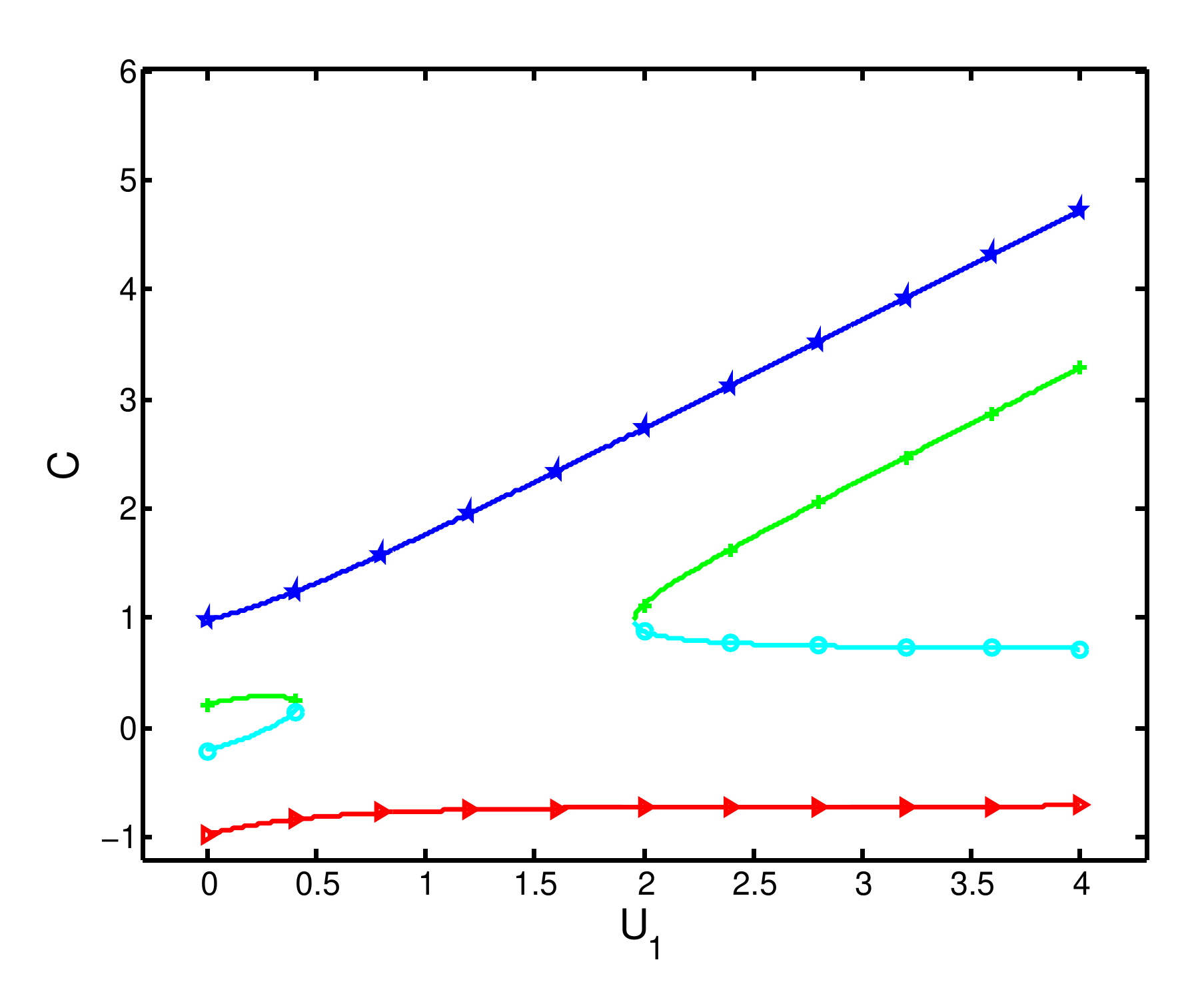} 
		\caption{Wave speeds (\ref{C}) as functions of the strength of the shear flow.}
    \end{figure}


\subsection{Critical layer}

The critical layer is a region in the neighbourhood of a line at which the local wave speed is equal to the shear flow speed, see  \cite{Freeman70, Johnson90, Johnson2012} and references therein. In this paper, we only consider a relatively weak shear flow, when  the critical layer does not appear, which we justify next.

The wavefront can be described as 
$$H(r,\theta,t)=rk(\theta)-st=\mbox{constant}.$$
Following \cite{Johnson90, Johnson_book},  we find that the local wave speed in the direction
$$\frac{\nabla H}{|\nabla H|}=\frac{k {\bf e_r}+k'{\bf e_\theta} }{\sqrt{k^2+k'^2}}$$ 
is given by
$$-\frac{H_t}{|\nabla H|}=\frac{s}{\sqrt{k^2+k'^2}}.$$
So the critical layer occurs when
$$\frac {s}{\sqrt{k^2+k'^2}}=(U_1-U_2)\cos(\alpha+\theta), \quad \mbox{where} ~\cos\alpha=\frac{k}{\sqrt{k^2+k'^2}},$$
which is equivalent to the condition
$$F_1=-s+(U_1-U_2)(k\cos\theta-k'\sin\theta)=0,$$
when the linear problem formulation fails. Note that $F_2 = -s \ne 0$.

We know that
$$F_{1\theta}=-(U_1-U_2)(k+k")\sin\theta,$$
where, without loss of generality, we assume that $U_1-U_2>0$ and $k+k">0$ on the selected singular solution ($k(\theta) > 0$). Then, $F_{1\theta}<0$ if $\theta\in (0,\pi)$ and $F_{1\theta}>0$ if $\theta\in(\pi,2\pi)$, which implies that $F_1$ reaches its maximum value at $\theta=0$. Therefore, to avoid the appearance of critical layers, we require that
$$
F_1 \le F_1|_{\theta = 0} = -s + (U_1 - U_2) k(0) < 0,
$$
which yields the following constraint on the strength of the shear flow:
\begin{equation}
(U_1 - U_2) k(0) < s.
\label{cl}
\end{equation}
We note that $k(0)$ depends on $U_1 - U_2$. However, since $s / k(\theta)$ represents the local wave speed in the direction of $\theta$, we know that $s/ k(0) \ge s$, implying that $k(0) \le 1$. Thus, we can replace the exact condition (\ref{cl}) with a simplified estimate:
\begin{equation}
U_1 - U_2 < s \le \frac{s}{k(0)}.
\label{crl}
\end{equation}

Thus, if the shear flow satisfies both conditions (\ref{ucl}) and (\ref{crl}), the long interfacial ring waves are K-H stable, and there are no critical layers. The coefficients of the derived $2+1$- dimensional amplitude equation (\ref{cKdV}) for both surface and interfacial ring waves in this two-layer case are listed in Appendix A.    

\section {Discussion}

In this paper, we developed an asymptotic theory describing the propagation of long linear and weakly-nonlinear ring waves in a stratified fluid over a shear flow  in the KdV regime. The theory is based on the existence of a suitable linear modal decomposition, which has more complicated structure than the known modal decomposition in Cartesian geometry, when waves propagate along or opposite the horizontal shear flow. In our formulation, the shear flow is horizontal, while the waves in the absence of the shear flow are concentric. Thus, there is a clash of geometries, and it is not clear a priori that there is any modal decomposition. 

The developed linear formulation provides, in particular,  a description of the distortion of the shape of the wavefronts of the ring waves by the shear flow, which has been illustrated by considering the classical setting of a two-layer fluid with a piecewise-constant current. The wavefronts of surface and interfacial ring waves were described in terms of two branches of the singular solution of the derived nonlinear first-order differential equation, constituting further generalisation of the well-known Burns and generalised Burns conditions \citep{Burns53, Johnson90}. Remarkably, the two branches of this singular solution could be described in parametric form, and an explicit analytical solution  was developed for the wavefront of the interfacial mode in the case of the low density contrast. The constructed solutions have revealed the qualitatively different behaviour of the wavefronts of surface and interfacial waves propagating over the same shear flow. Indeed,  while the wavefront of the surface ring wave is elongated in the direction of the flow, the wavefront of the interfacial wave is squeezed in this direction.

The derived $2+1$ - dimensional cylindrical Korteweg-de Vries type equation constitutes generalisation of the previously derived $1+1$ - dimensional equation for the surface waves in a homogeneous fluid over a shear flow \citep{Johnson90} and internal waves in a stratified fluid in the absence of a shear flow \citep{Lipovskii85}. Strictly speaking, Johnson has derived a $2+1$ - dimensional model \citep{Johnson90}, but as a by-product of our study we have shown that the complicated formula for one of the coefficients of his equation will yield zero coefficient for any shear flow, which then reduces the equation to a $1+1$ - dimensional model. Finally, for the case of the two-layer model we also derived a constraint on the strength of the shear flow, which guarantees that there are no critical layers,  and obtained explicit expressions for the coefficients of the derived amplitude equation in terms of the physical and geometrical parameters of the model, which provides a fully developed asymptotic theory for this case. Further work will include numerical and analytical studies of the long weakly - nonlinear ring waves using the derived equation.  

It should be noted that the derived model can be generalised to include the effects of variable environment and rotation, similar to the existing studies for the plane waves (see \cite{Ostrovsky, Grimshaw97, Grimshaw01, GH08, GHJ13} and references therein). The modal decomposition found in this paper can be used to derive other long ring wave models, similar to the Benjamin-Ono and intermediate-depth equations for the plane waves  \citep{Benjamin67, Ono, Joseph, Kubota}.

\begin{figure}
           \centering\includegraphics[width=0.8\columnwidth]{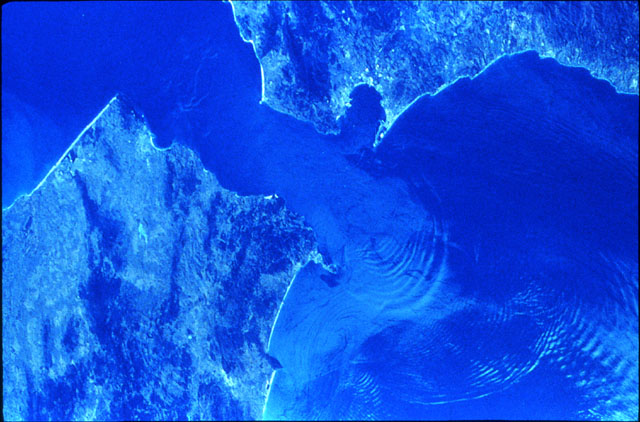} 
		\caption {Solitons in the Strait of Gibraltar (NASA image STS17-34-081, Lunar and Planetary Institute).}
\end{figure}

 Finally, we conjecture that theoretical results developed in this paper are relevant to the description of the nearly annular internal waves observed in the oceans. Such waves are generated in straits (e.g., in the Strait of Gibraltar, see Figure 1),  river - sea interaction zones (e.g., by a Columbia River plume, see \cite{Nash}) and by scattering from localised topographic features (e.g.,  from a sea mountain in the Celtic Sea, see \cite{Vlasenko3}). While the observed waves are often strongly-nonlinear, previous studies of such waves within the scope of the KdV-like models and their generalisations suggest that this is a useful asymptotic regime for this class of problems (see, for example, \cite{Helfrich06, Grimshaw97, GOSS98, Grue, Apel03, Apel07}). 

Squeezing of the wavefronts of interfacial ring waves in the direction of the  shear flow should be a prominent feature, and we conjecture that this might be a factor  contributing to the change of the shape of internal waves generated by an exchange flow in the Strait of Gibraltar, visible in Figure 11 for the waves propagating further into the Mediteranian Sea, as well as to the change of the shape of the internal spiral waves generated by a tidal flow near a sea mountain  in Figure A1 of \cite{Vlasenko3}. We hope that our study will help to better understand and interpret numerical and observational data for internal waves in  three - dimensional settings. 


\section{Acknowledgements}

We thank Ricardo Barros, Wooyoung Choi, Gennady El, Roger Grimshaw, John Grue, Robin Johnson,  Luigi Martina, Vladimir Matveev, Evgenii Kuznetsov, Alexander Mikhailov, Paul Milewski, Lev Ostrovsky, Victor Shrira, Yury Stepanyants, Vasyl Vlasenko and anonymous referees for useful references, criticisms and related discussions.
KK thanks team SKA from St. Petersburg, Russia for inspiration during the writing up stages of the first draft of this work. 
XZ thanks Xinhe Liu for useful discussions of some technical issues.


\appendix
\section{Coefficients of the cKdV-type equation}

In this Appendix we list the coefficients of the derived $2+1$- dimensional amplitude equation (\ref{cKdV}) for both surface and interfacial ring waves in the two-layer case. 

For the surface waves, we normalise $\phi$ by setting $\phi=1$ at $z=1$. The constant $\Lambda$ in the modal function (\ref{phi}) is given by
$$\Lambda_s=\frac{k^2+k'^2}{F_1^2}.$$ 
Substituting the modal function into the formulae (\ref{c1}) - (\ref{c5}), we obtain the coefficients in the form
{\small 
\begin{eqnarray*}
&\mu_1&=\frac{2s(k^2+k'^2)^2}{F_1^4} \left((1-d)\rho_1F_1+\frac{\rho_2F_2}{d}(\frac{F_1^2}{k^2+k'^2}+d-1)^2\right),\\
&\mu_2&=-\frac{3(k^2+k'^2)^3}{F_1^6}\left( (1-d)\rho_1F_1^2+\frac{\rho_2F_2^2}{d^2}(\frac{F_1^2}{k^2+k'^2}+d-1)^3\right),\\
&\mu_3&= -\frac{(k^2+k'^2)^3}{3F_1^4}\left(\rho_1F_1^2(\frac{F_1^6}{(k^2+k'^2)^3}-(\frac{F_1^2}{k^2+k'^2}+d-1)^3)+\rho_2F_2^2d(\frac{F_1^2}{k^2+k'^2}+d-1)^2\right),\\
&\mu_4&=-\frac{(1-d)\rho_1k(k+k")(k^2+k'^2)}{F_1^4}\left(F_1^2+4k'F_1(U_1-U_2)\sin\theta+3(k^2+k'^2)(U_1-U_2)^2\sin^2\theta\right)\\
&& -\frac{\rho_2(k+k")kF_2^2}{dF_1^4}(\frac{F_1^2}{k^2+k'^2}+d-1)\left((k^2-3k'^2)(\frac{F_1^2}{k^2+k'^2}+d-1)+\frac{4(d-1)k'(k'F_1+(U_1-U_2)(k^2+k'^2)\sin\theta)}{F_1}\right),\\
&\mu_5&=-\frac{2k(k^2+k'^2)}{F_1^4}\left((1-d)\rho_1F_1(k'F_1+(U_1-U_2)(k^2+k'^2)\sin\theta)+\frac{\rho_2k'F_2^2}{d}(\frac{F_1^2}{k^2+k'^2}+d-1)^2\right),
\end{eqnarray*} 
}
where 
\begin{eqnarray*}
F_1&=&-s+(U_1-U_2)(k\cos\theta-k'\sin\theta),\quad
F_2=-s,\\
s^2&=&\frac{1+\sqrt{(2d-1)^2+4\rho_1/\rho_2 d(1-d)}}{2},
\end{eqnarray*}
and the function $k(\theta)$ is defined by the formula (\ref{ss}) (lower sign).

For the interfacial waves, we normalise $\phi$ by setting $\phi=1$ at $z=d$. The constant $\Lambda$ in the modal function (\ref{phi}) is  given by
$$\Lambda_i=\frac{k^2+k'^2}{F_1^2+(d-1)(k^2+k'^2)}.$$ 
Substituting the modal function into the formulae (\ref{c1}) - (\ref{c5}), we obtain the coefficients in the form 

{\small
\begin{eqnarray*}
&\mu_1&=2s\left( \frac{(1-d)\rho_1F_1(k^2+k'^2)^2}{(F_1^2+(d-1)(k^2+k'^2))^2}+\frac{\rho_2F_2}{d}\right),\quad 
\mu_2=-3\left(\frac{(1-d)\rho_1F_1^2(k^2+k'^2)^3}{(F_1^2+(d-1)(k^2+k'^2))^3}+\frac{\rho_2F_2^2}{d^2}\right),\\
&\mu_3&=-\frac{\rho_1F_1^2}{3(F_1^2+(d-1)(k^2+k'^2))^2}\left(F_1^6-(F_1^2+(d-1)(k^2+k'^2))^3\right)-\frac 13 d \rho_2F_2^2(k^2+k'^2),\\
&\mu_4&=-\frac{(1-d)\rho_1k(k+k")\left((k^2-3k'^2)F_1^2-4k'(k^2+k'^2)F_1(U_1-U_2)\sin\theta-(U_1-U_2)^2(k^2+k'^2)^2\sin^2\theta\right)}{(F_1^2+(d-1)(k^2+k'^2))^2}\\
&&\qquad \qquad -\frac{4(1-d)\rho_1k(k+k")F_1^2}{(F_1^2+(d-1)(k^2+k'^2))^3}(k'F_1+(k^2+k'^2)(U_1-U_2)\sin\theta)^2-\frac{\rho_2k(k+k")(k^2-3k'^2)F_2^2}{{\color{red}d}(k^2+k'^2)^2},\\
&\mu_5&= -\frac{2(1-d)\rho_1F_1 k(k^2+k'^2)}{(F_1^2+(d-1)(k^2+k'^2))^2}(k'F_1+(U_1-U_2)(k^2+k'^2)\sin\theta)-\frac{2kk'\rho_2F_2^2}{d(k^2+k'^2)},
\end{eqnarray*}
}
where 
\begin{eqnarray*}
F_1&=&-s+(U_1-U_2)(k\cos\theta-k'\sin\theta),\quad
F_2=-s,\\
s^2&=&\frac{1-\sqrt{(2d-1)^2+4\rho_1/\rho_2 d(1-d)}}{2},
\end{eqnarray*}
and the function $k(\theta)$ is defined by the formula (\ref{ss}) (upper sign). For the latter, one can also use the explicit formula (\ref{Kovy}), obtained in the rigid lid approximation.



\begin{thebibliography}{99}



\bibitem[Ablowitz $\&$ Baldwin (2012)]{Ablowitz} Ablowitz, M.J. $\&$ Baldwin, D.E. 2012 Nonlinear shallow ocean wave soliton interactions on flat beaches. {\it Phys. Rev. E} {\bf 86}, 036305.

\bibitem[Alias {\it et al.} (2014)]{AGK} Alias, A., Grimshaw, R.H.J. $\&$ Khusnutdinova, K.R. 2014 Coupled Ostrovsky equations for internal waves in a shear flow. {\it Phys. Fluids} {\bf 26}, 126603.

\bibitem[Apel (2003)]{Apel03} Apel, J.R. 2003 A new analytical model for internal solitons in the ocean. {\it J. Phys. Oceanogr.} {\bf 33}, 2247 - 2269.

\bibitem[Apel {\it et al.}  (2007)]{Apel07}  Apel, J.R., Ostrovsky, L.A., Stepanyants, Y.A., Lynch, J.F. 2007 Internal solitons in the ocean and their effect on underwater sound.  {\it J. Acoust. Soc. Am.} {\bf 121}, 695 - 722.

\bibitem[Arkhipov {\it et al.} (2014)]{ASK} Arkhipov, D.G., Safarova, N.S., $\&$ Khabakhpashev, G.A. 2014 Dynamics of nonlinear three - dimensional waves on the interface between two fluids in a channel with low - sloping bottom and top. {\it Fluid Dynamics} {\bf 49}, 491 - 503.

\bibitem[Barros $\&$ Choi (2009)]{BC} Barros, R. $\&$ Choi, W. 2009 Inhibiting shear instability induced by large amplitude internal solitary waves in two - layer flows with a free surface. {\it Stud. Appl. Math.} {\bf 122}, 325 - 346.

\bibitem[Barros $\&$ Choi (2014)]{BarrosChoi} Barros, R. $\&$ Choi, W. 2014 Elementary stratified flows with stability at low Richardson Number. {\it Phys. Fluid} {\bf 26}, 124107.

\bibitem[Benjamin (1962)]{Benjamin62} Benjamin, T.B. 1962 The solitary wave on a stream with an arbitrary distribution of vorticity. {\it J. Fluid Mech.} {\bf 12}, 97 - 116.

\bibitem[Benjamin (1966)]{Benjamin66} Benjamin, T.B. 1966 Internal waves of finite amplitude and permanent form. {\it J. Fluid Mech.} {\bf 25}, 241 - 270.

\bibitem[Benjamin (1967)]{Benjamin67} Benjamin, T.B. 1967 Internal waves of permanent form in fluids of great depths. {\it J. Fluid Mech.} {\bf 29}, 559 - 592.

\bibitem[Benney (1966)]{Benney66} Benney, D.J. 1966 Long nonlinear waves in fluid flows. {\it J. Math. Phys.} {\bf 45}, 52 - 63.

\bibitem[Bontozoglou (1991)]{Bontozoglou91} Bontozoglou, V. 1991 Weakly nonlinear {Kelvin-Helmholtz} waves between fluids of finite depth. {\it Int. J. Multiphase Flow} {\bf 17}, 509 - 518.

\bibitem[Boonkasame $\&$ Milewski (2011)]{BM} Boonkasame, A. $\&$ Milewski, P.A. 2011 The stability of large-amplitude shallow interfacial {non-Boussinesq} flows. {\it Stud. Appl. Math.} {\bf 128}, 40 - 58.

\bibitem[Boonkasame $\&$ Milewski (2014)]{Boonkasame} Boonkasame, A. $\&$ Milewski, P.A. 2014 A model for strongly nonlinear long interfacial waves with background shear. {\it Stud. Appl. Math.} {\bf 133}, 182 - 213.

\bibitem[Boussinesq (1871)]{Boussinesq} Boussinesq, J. 1871 Th\'eorie de l'intumescence liquide appel\'ee onde solitaire ou de translation, se propageant dans un canal rectangulaire. {\it Comptes Rendus Acad. Sci. (Paris)} {\bf 72}, 755 - 759.

\bibitem[Buhler (2009)]{Buhler} Buhler, O. 2009 {\it Waves and mean flows}. Cambridge: Cambridge University Press.

\bibitem[Burns (1953)]{Burns53} Burns, J.C. 1953  Long waves in running water. {\it Proc. Camb. Phil. Soc.} {\bf 49}, 695 - 706.

\bibitem[Calogero $\&$ Degasperis (1978)]{Calogero} Calogero, F. $\&$ Degasperis, A. 1978 Solution by the spectral transform method of a nonlinear evolution equation including as a special case the cylindrical KdV equation. {\it Lett. Nuovo Cim.} {\bf 23}, 150 - 154.

\bibitem[Chakravarty $\&$ Kodama (2014)]{Kodama} Chakravarty, S. $\&$ Kodama, Y. 2014 Construction of {KP} solitons from wave patterns. {\it J. Phys. A} {\bf 47}, 025201.

\bibitem[Choi (2006)]{Choi} Choi, W. 2006 The effect of a background shear current on large amplitude internal solitary waves. {\it Phys. Fluids} {\bf 24}, 1 - 7.

\bibitem[Choi $\&$ Camassa (1999)]{CC} Choi, W. $\&$ Camassa, R. 1999 Fully nonlinear internal waves in a two - fluid system. {\it J. Fluid Mech.} {\bf 396}, 1 - 36.

\bibitem[Chumakova {\it et al.} (2009)]{Chumakova} Chumakova, L.,  Menzaque F.E.,  Milewski, P.A.,  Rosales, R.R.,  Tabak, E.G., $\&$ Turner, C.V. 2009 Stability properties and nonlinear mappings of two and three-layer stratified flows. {\it Stud. Appl. Math.} {\bf 122}, 123 - 137.

\bibitem[Constantin {\it et al.} (2015)]{Constantin} Constantin, A., Kalimeris, K., Scherzer, O. 2015 Approximations of steady periodic water waves in flows with constant vorticity. {\it Nonlin. Anal.: Real World Appl.} {\bf 25}, 276 - 306.




\bibitem[Craik (1985)]{Craik} Craik, A.D.D. 1985 {\it Wave interactions and fluid flows.} Cambridge: Cambridge University Press.

\bibitem[Drazin $\&$ Reed (2004)]{Drazin} Drazin, P.G. $\&$ Reed, W.H. 2004 {\it Hydrodynamic stability. } Cambridge: Cambridge University Press.


\bibitem[Ellingsen (2014)]{Ellingsen} Ellingsen, S.A. 2014 Ship waves in the presence of uniform vorticity. {\it J. Fluid Mech.} {\bf 742}, R2-1 - R2-11.

\bibitem[Farmer $\&$ Armi (1988)]{Farmer} Farmer, D.M. $\&$ Armi, L. 1988 The flow of Atlantic water through the Strait of Gibraltar. {\it Prog. Oceanogr.} {\bf 21}, 1 - 105.

\bibitem[Freeman $\&$ Johnson (1970)]{Freeman70} Freeman, N.C. $\&$ Johnson, R.S. 1970 Shallow water waves on shear flows. {\it J. Fluid Mech.} {\bf 42}, 401 - 409.

\bibitem[Grimshaw {\it et al.} (1997)]{Grimshaw97} Grimshaw, R.H.J., Pelinovsky, E. $\&$ Talipova, T. 1997 The modified {Korteweg - de Vries} equation in the theory of large-amplitude internal waves. {\it Nonlin. Processes Geophys.} {\bf 4}, 237 - 250.

\bibitem[Grimshaw {\it et al.} (1998)]{GOSS98}  Grimshaw, R.H.J., Ostrovsky, L.A., Shrira, V.I. $\&$ Stepanyants, Yu. A. 1998 Long nonlinear surface and internal gravity waves in a rotating ocean. {\it Surveys in Geophysics} {\bf 19}, 289 - 338.

\bibitem[Grimshaw (2001)]{Grimshaw01} Grimshaw, R.H.J. 2001 Internal solitary waves. In {\it Environmental Stratified Flows} (ed. R. Grimshaw), pp. 1 - 27. Kluwer.

\bibitem[Grimshaw $\&$ Helfrich (2012)]{GH08} Grimshaw, R. $\&$ Helfrich, K.R. 2012 The effect of rotation on internal solitary waves. {\it IMA J. Appl. Math.} {\bf 77}, 326 - 339. 

\bibitem[Grimshaw {\it et al.} (2013)]{GHJ13} Grimshaw, R., Helfrich, K. $\&$ Johnson, E. 2013 Experimental study of the effect of rotation on large amplitude internal waves. {\it Phys. Fluids} {\bf 25}, 056602.

\bibitem[Grue (2006)]{Grue}  Grue, J. 2006 Very large internal waves in the ocean - observations and nonlinear models. In {\it Waves in Geophysical Fluids} (ed. J. Grue $\&$ K. Trulsen), pp. 1 - 66. Springer.

\bibitem[Grue (2015)]{Grue15} Grue, J. 2015 Nonlinear interfacial wave formation in three dimensions. {\it J. Fluid Mech.} {\bf 767}, 735-762.

\bibitem[Helfrich $\&$ Melville (2006)]{Helfrich06} Helfrich, K.R. $\&$ Melville, W.K. 2006 Long nonlinear internal waves. {\it Ann. Rev. Fluid Mech.} {\bf 38}, 395 - 425.

\bibitem[Jackson {\it et al.} (2013)]{Jackson} Jackson, C.R., Da Silva, J.C., Jeans, G., Alpers, W. $\&$ Caruso, M.J. 2013 Nonlinear internal waves in synthetic aperture radar imagery. {\it Oceanography} {\bf 26}, 68-79.

\bibitem[Johnson (1980)]{Johnson80} Johnson, R.S. 1980 Water waves and Korteweg - de Vries equations. {\it J. Fluid Mech.} {\bf 97}, 701 - 719.

\bibitem[Johnson (1990)]{Johnson90} Johnson, R.S. 1990 Ring waves on the surface of shear flows: a linear and nonlinear theory. {\it J. Fluid Mech.} {\bf 215}, 145 - 160.

\bibitem[Johnson (1997)]{Johnson_book} Johnson, R.S. 1997 {\it A modern introduction to the mathematical theory of water waves. } Cambridge: Cambridge University Press.

\bibitem[Johnson (2012)]{Johnson2012} Johnson, R.S. 2012 Models for the formation of a critical layer in water wave propagation. {\it Phil. Trans. R. Soc.} {\bf 370}, 1638 - 1660.

\bibitem[Joseph (1977)]{Joseph} Joseph, R.I. 1977 Solitary waves in a finite depth fluid. {\it J. Phys. A Math. Gen.} {\bf 10}, L1225 - L1227.

\bibitem[Klein {\it et al.} (2007)]{Klein07} Klein, C., Matveev, V.B. $\&$ Smirnov, A.O. 2007 The cylindrical {Kadomtsev - Petviashvili} equation: old and new results. {\it Theor. Math. Phys.} {\bf 152}, 1132 - 1145.

\bibitem[Korteweg $\&$ de Vries (1895)]{KdV} Korteweg, D.J. $\&$ de Vries, G. 1895 On the change of form of long waves advancing in a rectangular channel, and on a new type of long stationary waves. {\it Philos. Mag.} {\bf 39}, 422 - 443.

\bibitem[Kubota {\it et al.} (1978)]{Kubota} Kubota, T., Ko, D.R., $\&$ Dobbs, L. 1978 Weakly - nonlinear long internal waves in a stratified fluid of finite depth. {\it J. Hydronautics} {\bf 12}, 157 - 165.

\bibitem[Lannes $\&$ Ming (2015)]{Lannes} Lannes, D. $\&$ Ming, M. 2015 The Kelvin - Helmholtz instabilities in two-fluids shallow water models. {\it to appear in Communication of the Fields Institute. $<hal-01101993>$}

\bibitem[Lee $\&$ Beardsley (1974)]{Lee}  Lee, C.-Y., Beardsley, R.C. 1974 The generation of long nonlinear internal waves in a weakly stratified shear flow. {\it J. Geophys. Res.} {\bf 79}, 453 - 462.

\bibitem[Lipovskii (1985)]{Lipovskii85} Lipovskii, V.D. 1985 On the nonlinear internal wave theory in fluid of finite depth. {\it Izv. Akad. Nauk SSSR, Ser. Fiz. Atm. Okeana} {\bf 21}, 864 - 871.

\bibitem[Long (1955)]{Long55} Long, R.R. 1955 Long waves in a two-fluid system. {\it J. Meteorol.} {\bf 13}, 70 - 74.

\bibitem[Maslowe $\&$ Redekopp (1980)]{MR} Maslowe, S.A. $\&$ Redekopp, L.G. Long nonlinear waves in stratified shear flows. {\it J. Fluid Mech.} {\bf 101}, 321 - 348.

\bibitem[Maxon $\&$ Viecelli (1974)]{Maxon74} Maxon, S. $\&$ Viecelli, J. 1974 Spherical solitons. {\it Phys. Rev. Lett.} {\bf 32}, 4 - 6.

\bibitem[Miyata (1985)]{Miyata85} Miyata, M. 1985 An internal solitary wave of large amplitude. {\it La Mer.} {\bf 23}, 43 - 48.

\bibitem[Miles (1978)]{Miles78} Miles, J.W. 1978 An axisymmetric Boussinesq wave. {\it J. Fluid Mech.} {\bf 84}, 181 - 191.

\bibitem[Mooers (1975)]{Mooers} Mooers, C.N.K. 1975 Several effects of a baroclinic current on the cross-stream propagation of inertial internal waves. {\it Geophys. Fluid Dyn.} {\bf 6}, 245 - 475.

\bibitem[Nash $\&$ Moum (2005)]{Nash} Nash, J.D. $\&$ Moum, J.N. 2005 River plums as a source of large-amplitude internal waves in the coastal ocean. {\it Nature} {\bf 437}, 400-403.

\bibitem[Nwogu (2009)]{Nwogu} Nwogu, O.G. 2009 Interaction of finite-amplitude waves with vertically sheared current fields. {\it J. Fluid Mech.} {\bf 627}, 179 - 213.

\bibitem[Oikawa {\it et al.} (1987)]{Oikawa} Oikawa, M., Chow, K., Benney, D.J.,  1987 The propagation of nonlinear wave packets in a shear flow with a free surface. {\it Stud. Appl. Math,} {\bf 76}, 69 - 92.

\bibitem[Olbers (1981)]{Olbers} Olbers, D.J. 1981 The propagation of internal waves in a geostrophic current. {\it J. Phys. Oceanogr.} {\bf 11}, 1224 - 1233.

\bibitem[Ono (1975)]{Ono} Ono, H. Algebraic solitary waves in stratified fluid. {\it J. Phys. Soc. Jpn.} {\bf 39}, 1082 - 1091.

\bibitem[Ovsyannikov (1979)]{Ovsyannikov} Ovsyannikov, L.V. 1979 Two-layer "shallow water" model. {\it J. Appl. Meth. Tech. Phys.,} {\bf 20}, 127-135.

\bibitem[Ovsyannikov (1985)]{Ovsyannikov_book} Ovsyannikov, L.V. 1985  {\it Nonlinear problems in the theory of surface and internal waves.} Moscow: Nauka (in Russian).

\bibitem[Ostrovsky (1978)]{Ostrovsky} Ostrovsky, L. 1978 Nonlinear internal waves in a rotating ocean. {\it Oceanology,} {\bf 18}, 119 - 125.

\bibitem[Ramirez {\it et al.} (2002)]{Ramirez} Ramirez, C., Renouard, D.,  Stepanyants, Yu.A. 2002 Propagation of cylindrical waves in a rotating fluid. {\it Fluid Dyn. Res.} {\bf 30}, 169 - 196.

\bibitem[Sannino {\it et al.} (2014)]{Sannino} Sannino, G., Sanchez Garrido, J.C., Liberti, L. $\&$ Pratt, L. 2014 Exchange flow through the Strait of Gibraltar as simulated by a $\sigma$-coordinate hydrostatic model and a $z$-coordinate nonhydrostatic model. In {\it The Mediterranean Sea: Temporal Variability and Spatial Patterns} (ed. Gianluca Eusebi Borzelli, Miroslav Gacic, Piero Lionello, Paola Malanotte-Rizzoli), AGU Book - Wiley.


\bibitem[Stastna $\&$ Lamb (2002)]{Stastna1}  Stastna, M., Lamb, K.G.  2002 Large fully nonlinear solitary waves: the effect of background current. {\it Phys. Fluids} {\bf 14}, 2987.

\bibitem[Stastna $\&$ Walter (2014)]{Stastna2}  Stastna, M., Walter, R. 2014 Transcritical generation of nonlinear internal waves in the presence of background shear flow. {\it Phys. Fluids} {\bf 26}, 086601.

\bibitem[Thomas {\it et al.} (2012)]{Thomas}  Thomas, R., Kharif, C., Manna, M.  2012 A nonlinear {Schr$\ddot o$dinger} equation for water waves on finite depth with constant vorticity. {\it Phys. Fluids} {\bf 24}, 127102.

\bibitem[Turner (1973)]{Turner} Turner, J.S. {\it Buoyancy effects in fluids.} Cambridge: Cambridge University Press.


\bibitem[Vlasenko {\it et al.} (2009)]{Vlasenko1} Vlasenko, V., Sanchez Garrido, J.C., Stashchuk, N., Garcia Lafuente, J., Losada, M. 2009 Three-dimensional evolution of large-amplitude internal waves in the Strait of Gibraltar. {\it J. Phys. Oceanogr.} {\bf 39}, 2230 - 2246.


\bibitem[Vlasenko {\it et al.} (2013)]{Vlasenko2}  Vlasenko, V., Stashchuk, N., Palmer, M.R., Inall, M.E., 2013 Generation of baroclinic tides over an isolated underwater bank.  {\it J. Geophys. Res.} {\bf 118}, 4395 - 4408.

\bibitem[Vlasenko {\it et al.} (2014)]{Vlasenko3}   Vlasenko, V., Stashchuk, N., Inall M.E., Hopkins, J.E., 2014 Tidal energy conversion in a global hot spot: on the 3-D dynamics of baroclinic tides at the Celtic Sea shelf break. {\it Phys. Fluids} {\bf 24}, 127102.

\bibitem[Voronovich {\it et al.} (2006)]{Voronovich} Voronovich, V.V., Sazonov, I.A. $\&$ Shrira, V.I. 2006 On radiating solitons in a model of the internal wave - shear flow resonance. {\it J. Fluid Mech.} {\bf 568}, 273 - 301.



\bibitem[Weidman $\&$ Zakhem (1988)]{Weidman} Weidman, P.D. $\&$ Zakhem, R. 1988 Cylindrical solitary waves. {\it J. Fluid Mech.} {\bf 191}, 557 - 573.

\bibitem[Young {\it et al.} (1982)]{Young}   Young, W.R., Rhines, P.B., Garrett, C.J.R. 1982 Shear-flow dispersion, internal waves and horizontal mixing in the ocean. {\it J. Phys. Oceanogr.} {\bf 12}, 515 - 527.









\end{thebibliography}

\end{document}